\documentclass[epj]{svjour}
   \usepackage{graphics}
   \usepackage{graphicx}
   \usepackage{multirow}
   \usepackage{colortbl}
   \usepackage{verbatim}
   \usepackage{amsmath}
   \usepackage{enumerate}
   \usepackage{amssymb}
  
\begin{document}
\title{Design Studies of the PWO Forward End-cap Calorimeter for $\overline{\rm{P}}$ANDA}
\author{H. Moeini\inst{1}
\thanks{\emph{Corresponding author:} moeini@kvi.nl}
\and M. Al-Turany\inst{2}
\and M. Babai\inst{1}
\and A. Biegun\inst{1}
\and O. Bondarenko\inst{1}
\and K. G\"{o}tzen\inst{2}
\and M. Kavatsyuk\inst{1}
\and M.F.~Lindemulder\inst{1} 
\and H. L\"{o}hner\inst{1} 
\and D. Melnychuk\inst{3} 
\and J.G. Messchendorp\inst{1} 
\and H.A.J. Smit\inst{1} 
\and S. Spataro\inst{4}
\and R. Veenstra\inst{1}, for the $\overline{\rm{P}}$ANDA collaboration
\thanks{http://www-panda.gsi.de}
}                     
%
%
\institute
{Kernfysisch Versneller Instituut (KVI), University of Groningen, Groningen, The Netherlands 
 \and GSI Helmholtzzentrum f\"{u}r Schwerionenforschung GmbH 
 \and National Centre for Nuclear Research, Warsaw, Poland 
 \and Dipartimento di Fisica, Universit\`{a} di Torino and INFN, Italy
}
\date{Received: date / Revised version: date}
%
\abstract{
The $\overline{\rm{P}}$ANDA detection system at FAIR, Germany, is designed to study antiproton-proton annihilations, in order to investigate among others the 
realm of charm-meson states and glueballs, which has still much to reveal. The yet unknown properties of this field are to be unraveled through studying QCD phenomena in the non-perturbative regime. 
The multipurpose $\overline{\rm{P}}$ANDA detector will be capable of tracking, calorimetry, and particle identification, and is foreseen to run at high luminosities providing average 
reaction rates up to $2\cdot10^{7}$~interactions/s. The envisaged physics program requires measurements of photons and charged particles with excellent energy, position, and time 
resolutions. The electromagnetic calorimeter (EMC) will serve as one of the basic components of the detector setup and comprises cooled Lead-Tungstate (PbWO$_4$) crystals.
This paper presents the mechanical design of the Forward End-cap calorimeter and analyzes the response\thanks{The intention of this paper is to study constraints imposed by the mechanical 
design rather than giving a description of the complete technical design and analysis of the Forward End-cap.} of the Forward End-cap calorimeter in conjunction with the full EMC and 
the complete $\overline{\rm{P}}$ANDA detector.
The simulation studies are focused on the performance of the planned EMC with respect to the energy and spatial resolution of the reconstructed photons.
Results of the Monte Carlo simulations, excluding very low-energy photons, have been validated by data obtained from a prototype calorimeter and shown to fulfil 
the requirements imposed by the $\overline{\rm{P}}$ANDA physics program.
%
\keywords{charmonium, electromagnetic calorimeter, electromagnetic showers, energy resolution, full-energy efficiency, Monte Carlo simulations}
} 
\maketitle
\section{Introduction}
\label{Introduction_label}
The $\overline{\rm{P}}$ANDA Experiment \cite{PhysicsBook} will be one of the key experiments in hadron physics at the Facility for Antiproton and Ion Research (FAIR) at Darmstadt, Germany. The 
central part of FAIR is a synchrotron complex providing intense pulsed ion beams ranging from protons to Uranium. Antiprotons with a momentum between 1.5~GeV/c and 15~GeV/c, 
produced by a primary proton beam, will be stored and cooled in the High Energy Storage Ring (HESR) and interact in a fixed target geometry in the $\overline{\rm{P}}$ANDA detector. In the 
high resolution mode, a RMS momentum resolution $\frac{\sigma_{p}}{p}\leq4\cdot10^{-5}$ is envisaged for the beam momenta between 1.5~GeV/c and 8.9~GeV/c.
The average peak luminosity of $2\cdot10^{31}~\rm{cm}^{-2}\cdot\rm{s}^{-1}$ for $10^{10}~\overline{\rm{p}}$ is expected for this mode, 
assuming $\rho_{target}=4\cdot10^{15}~\rm{atoms}\cdot\rm{cm}^{-2}$.
With the same target density and for $10^{11}~\overline{\rm{p}}$, an average peak luminosity of $2\cdot10^{32}~\rm{cm}^{-2}\cdot\rm{s}^{-1}$ is expected for the high luminosity mode, providing a RMS momentum spread of $\frac{\sigma_{p}}{p}\approx10^{-4}$ over 
the momentum range of 1.5~GeV/c to 15~GeV/c \cite{Lehrach}. A hydrogen cluster jet or a hydrogen pellet target will allow an average interaction rate of up to 20~MHz. 
Contrary to $\rm{e}^{+}\rm{e}^{-}$ collisions, states of all non-exotic quantum numbers can be formed directly in antiproton-proton annihilations.
This allows mass and width measurements of hadronic resonances by the beam-scanning technique \cite{PhysicsBook,PandaTechDesRep} with an accuracy of $50-100$~keV, which is 10 to 100 times 
better than realized in any $\rm{e}^{+}\rm{e}^{-}$-collider experiment. 

\begin{figure}
\begin{center}
\resizebox{0.48\textwidth}{!}{\includegraphics{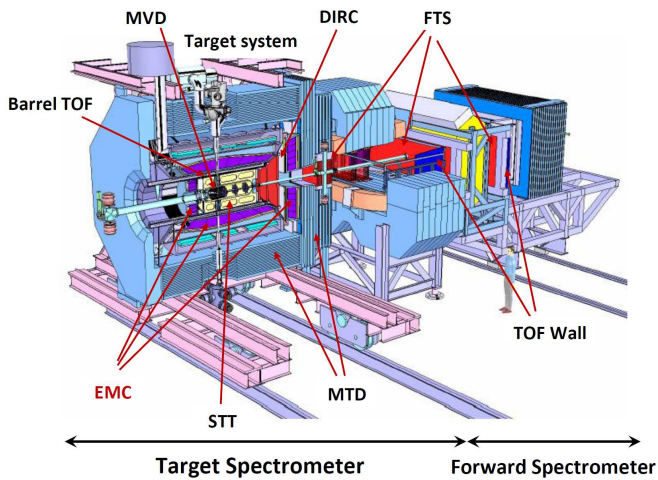}}
\caption{Schematic view of the $\overline{\rm{P}}$ANDA detector consisting of Target Spectrometer (located inside the solenoid magnet) and Forward Spectrometer. The EMC, as part of 
the target spectrometer, and other detector elements are labeled in the Figure with explanations given in the text. The scale is indicated by the person in the Figure.}
\label{panda_full}
\end{center}
\end{figure}
\begin{sloppypar}
In order to collect all the relevant kinematic information from the final states of the antiproton-proton collisions, the experiment employs the versatile $\overline{\rm{P}}$ANDA detector which is able to 
provide precise trajectory reconstruction, measure energy and momentum of particles with high resolution, and efficiently identify the charged particles. The detector
(Fig.~\ref{panda_full}) is subdivided into two magnetic spectrometers: the Target Spectrometer (TS), located inside a 2~T superconducting solenoid magnet surrounding the interaction point and  
covering angles from $5^{\circ}$($10^{\circ}$) to $170^{\circ}$ in the vertical (horizontal) plane; the Forward Spectrometer (FS), based on a dipole magnet with a field integral 
of up to 2~Tm to momentum-analyze forward-scattered particles in the region of angles up to $5^{\circ}$ ($10^{\circ}$). 
The combination of the two spectrometers allows for tracking, momentum reconstruction, charged-particle identification, and electromagnetic calorimetry as well as muon 
identification in a close to $4\pi$ geometry. The various elements of the $\overline{\rm{P}}$ANDA detector system include the Micro Vertex detector (MVD), Straw 
Tube Tracker (STT), Gas Electron Multiplier (GEM), Cherenkov detectors (DIRC: Detection of Internally Reflected Cherenkov light), Forward Tracking Stations (FTS), Muon Tracking 
Detector (MTD), Aerogel Ring Imaging Cherenkov Counter (RICH), Time Of Flight System (SciTil), the Electromagnetic Calorimeter (EMC), and the Forward Time-Of-Flight Wall (FTOF).
\end{sloppypar}
\par The experiment focuses on hadron spectroscopy, in particular on the search for exotic states in the charmonium mass region, on the interaction of charmed hadrons with the 
nuclear medium, on double-hypernuclei to investigate the nuclear potential and hyperon-hyperon interactions as well as on electromagnetic processes to study various aspects of 
nucleon structure~\cite{PandaTechDesRep}. These physics objectives define the requirements for the $\overline{\rm{P}}$ANDA detector system in which the EMC plays a crucial role. 
For precision spectroscopy of charmonium 
states and exotic hadrons in the charmonium region a full acceptance is required to allow for a proper partial-wave analysis. Accordingly, as final states with many photons can occur, 
a low photon threshold of about 10~MeV is a key requirement for the electromagnetic calorimeter. Consequently, a $2-3$~MeV threshold for individual crystals and low noise levels 
of about 1~MeV are required.
\par This paper demonstrates the performance of one of the important modules of the EMC, the Forward End-cap calorimeter in conjunction with other EMC modules.
To this end the mechanical design, as implemented in the Monte Carlo simulation tool, will be explained in Section \ref{EMCofPANDA_label}.
The analysis that is presented throughout the paper will not be applicable for very low-energy photons.
The results of the Monte Carlo simulations, based on experimental studies of the prototype, will 
be discussed in Section \ref{Simulations_label}, and the performance of the detection system for the benchmark physics channel 
$\overline{p}+{p} \rightarrow h_{c} \rightarrow \eta_{c}+\gamma \rightarrow (\pi^{0}+\pi^{0}+\eta)+\gamma \rightarrow 7\gamma$ will be presented in Section \ref{Charmonium_label}. Although all 
the EMC modules (including Forward End-cap, Backward End-cap, and Barrel) will be employed for the complete reconstruction of this decay channel, the optimized parameters for the energy 
reconstruction that are obtained from a prototype setup made of Barrel-type crystals will be exploited for the Barrel crystals. Such an energy optimization is required for Barrel crystals
due to light yield non-uniformities (see Section~\ref{Light-yield-non-uniformity}) which appear to be absent in the End-cap crystals.
Simulation results, excluding very low-energy photons, will show that the proposed technical design of the calorimeter meets the requirements imposed by the $\overline{\rm{P}}$ANDA physics program.

\section{The Electromagnetic Calorimeter of $\overline{\rm{P}}$ANDA}
\label{EMCofPANDA_label}
\subsection{Operating conditions and layout}
\label{EMClayout_label}
In the TS, high precision electromagnetic calorimetry is required over a large energy range from a few MeV up to 15~GeV. Lead-tungstate (PbWO$_4$ or briefly PWO) is chosen as 
calorimeter material in the TS due to its fast response and correspondingly high count-rate capability, and its high density which allows for a compact setup.
Despite the low light yield, PWO revealed good energy resolution for photon and electron detection at intermediate energies~\cite{Rainer_1}, which motivates the detailed 
response study presented here.
The high granularity and maximum acceptance of the EMC are needed, in order to discriminate pions from electrons for momenta above 0.5~GeV/c.
In high-energy physics PWO has been chosen by the CMS collaboration at CERN \cite{CMS_TechnicalProposal}. 
For $\overline{\rm{P}}$ANDA, it is proposed to use crystals with a length of 200~mm ($\approx22$ radiation lengths), which allows optimum shower 
containment for photons up to 15~GeV and limits the nuclear counter effect in the subsequent photo-sensor to a tolerable level, in order to achieve an energy resolution ($\sigma/E$) for photons 
and electrons of $1.95(4)\% / \sqrt{E [\rm{GeV}]}+0.48(5)\%$ \cite{Kavatsyuk}. All crystals will be cooled down to $-25^{\circ}$~C to provide a light yield of 500 photons/MeV which constitutes an overall 
gain factor of 3.5 as compared to the operation at $+18^{\circ}$~C. Since $\overline{\rm{P}}$ANDA is a fixed-target experiment, produced particles are boosted in the forward direction and 
the event-rate distribution in the EMC is not isotropic. An average event rate of 10 to 100~kHz is expected in the central part around the interaction region covered by the Barrel EMC, 
whereas the forward region covered by the Forward End-cap EMC (FwEndCap) will be exposed to event rates of up to 500~kHz. Thus, two large-area avalanche photo-diodes (LAAPD) \cite{PandaTechDesRep}, 
each having a 100~mm$^{2}$ area, will be used as photo-sensors for the Backward End-cap EMC (BwEndCap), the Barrel, and the outermost 80\% crystals of the FwEndCap. The innermost 20\% crystals 
of the FwEndCap will be equipped with vacuum photo-tetrodes (VPTT) \cite{PandaTechDesRep}, in order to adapt to the expected extreme high rates of up to 500 kHz in the 
region nearest to the beam axis.
\begin{figure}
\begin{center}
\resizebox{0.48\textwidth}{!}{\includegraphics{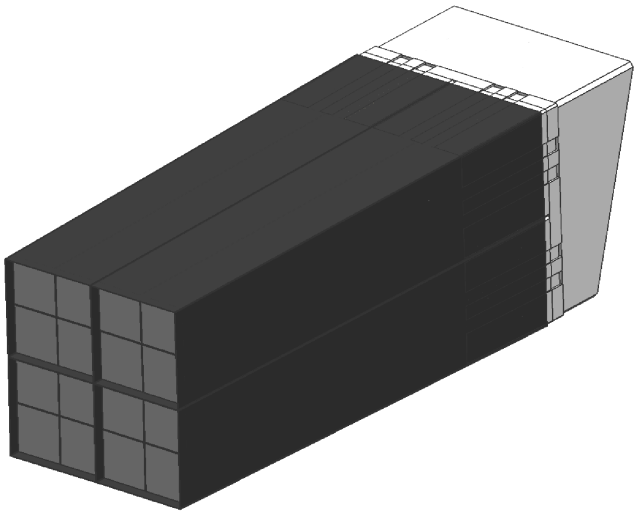}}
\caption{Subunit of 16 crystals assembled in a Carbon-fiber alveole with mounting interface.}
\label{subunit}
\end{center}
\end{figure}
\par The $\overline{\rm{P}}$ANDA PWO crystals will be arranged in the cylindrical Barrel with a length of 2.5~m and an inner radius of about 57~cm (11360 crystals), in the BwEndCap 
(528 crystals), and in the FwEndCap (3856 crystals). These three separate parts of the EMC, referred to as EMC modules in the target spectrometer, will be placed 
inside the 2~T solenoid magnet of the TS. The basic crystal shapes are squared-profile prismoids with two of their neighboring sides being right-angled trapezoids, based on 
the ``flat-pack" configuration used in the CMS calorimeter~\cite{CMS_TechDesRep}. The dimensions of the crystals are about $20\times20\times200$~mm$^3$ for the Barrel and BwEndCap. 
The same crystal length of 200~mm is used for the FwEndCap, but with slightly larger crystal areas of $24.37\times24.37$~mm$^2$ at the front and $26\times26$~mm$^2$ at the rear faces.

\subsection{The FwEndCap verification strategy}
\label{FwEndCap_label}
The preliminary concepts related to the EMC detector are detailed in the Technical Design Report~\cite{PandaTechDesRep}. The FwEndCap design has to fulfil the demands imposed by the 
physics program of $\overline{\rm{P}}$ANDA. This requires measurements of photons and charged particles with a good energy, time, and spatial resolution. It is extremely important 
that the FwEndCap can take the high count rates with a sufficient resolution without being damaged. The complete coverage of the forward region is necessary for the most efficient 
reconstruction of all reaction products.
\par Inspired by the CMS design \cite{CMS_TechnicalProposal}, the mechanical layout of the FwEndCap was designed according to the requirements and geometrical restrictions of the 
$\overline{\rm{P}}$ANDA detector. In connection with the development of the computing framework PandaROOT~\cite{ref_PandaROOT} for $\overline{\rm{P}}$ANDA, the verification of the mechanical 
design through simulations and a validation with experimental results is mandatory. PandaROOT is a framework for both simulations and data analysis, 
and is mainly based on the object-oriented data analysis framework ROOT~\cite{ROOT}. It features the concept of Virtual Monte Carlo~\cite{VirtualMC}, which allows to run transport 
models {\tt Geant3} and {\tt Geant4}~\cite{GEANT} using the same code, making it easy to compare the results of various transport models with exactly the same conditions. Using PandaROOT the designed 
geometry of the FwEndCap (crystals and Carbon Fibre packages) was used for detector simulation. The simulation results will be discussed in the remainder of this article. Here, all the 
simulation results are obtained using the {\tt Geant3} transport model and a production cut energy of 1~MeV is imposed for secondary particles.
\begin{figure}
\begin{center}
\resizebox{0.48\textwidth}{!}{\includegraphics{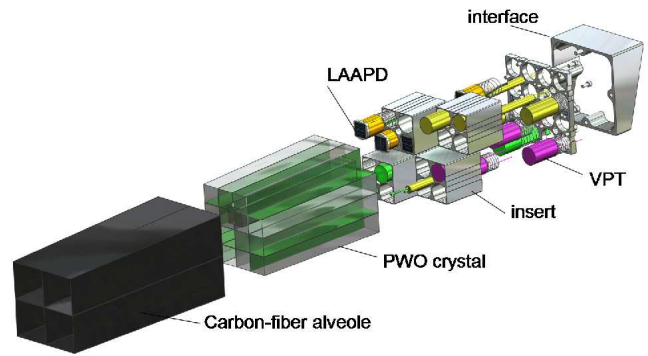}}
\caption{Exploded-view of one subunit with (looking from the left): Carbon-fiber alveole separating the boxes in a subunit, 16 crystals of a subunit, inserts holding either LAAPD or VPTT 
photo-sensors, and the mounting interface. Only one sort of photo-sensor is used in a subunit.
}
\label{subunit-explosion}
\end{center}
\end{figure}
\begin{figure*}
\begin{center}
\resizebox{0.8\textwidth}{!}{\includegraphics{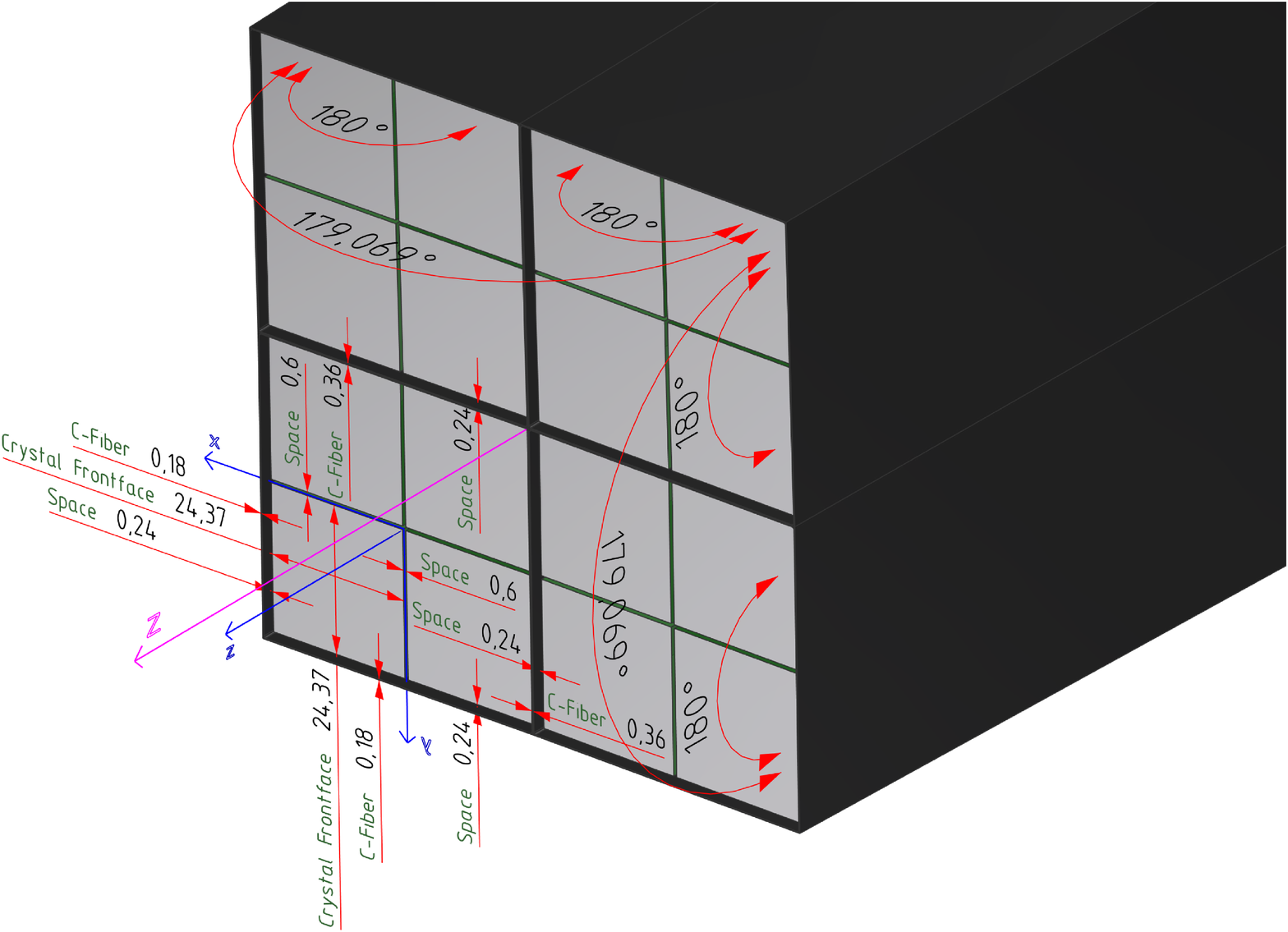}}
\caption{Dimensions of the subunit construction (in mm). Groups of four crystals are arranged in a box with a planar front face. The front faces of either two boxes are slightly tilted with a 
relative angle of ca. 1$^{\circ}$. The central axes of a box ({\bf z}) and of the subunit ({\bf Z}) are indicated. The width of a crystal and the thickness of the Carbon-fiber walls 
and gaps are shown.}
\label{subunit-dimensions}
\end{center}
\end{figure*}

\subsection{The FwEndCap technical design}
\label{FwEndCapDesign_label}
Based on the technical design of the FwEndCap, 16 crystals will be composed in packages called subunits (see Fig.~\ref{subunit}). 
A schematic representation of the various components forming a subunit is shown in Fig.~\ref{subunit-explosion}.
The crystals of a subunit are contained by a layer of 0.18~mm Carbon-fibre material (see Fig.~\ref{subunit-dimensions}) forming the walls of a container called {\it alveole} which was 
produced by Fiberworx BV~\cite{fiberworkx} according to our design specifications. Two side-by-side boxes in the geometry of the FwEndCap are separated by a 0.36~mm thick alveole wall. 

In the simulations, every four crystals are arranged in a box of $2\times2$ crystals, 
which is a squared-profile right prismoid or, technically, a frustum (truncated pyramid). The specific shape of the crystals ensures that all the front and 
rear faces of the crystals in a box lie on two parallel planes 200~mm apart. Every subunit comprises four identical boxes whose symmetry axes are slightly rotated with respect to the subunit 
symmetry axis ({\bf Z} in Fig.~\ref{subunit-dimensions}). The boxes are rotated around the axes {\bf x} and {\bf y} of a box by the same angle   
($\Delta\theta_{y} = \Delta\theta_{x} = \rm{atan}[24.37/3000]$) in such a way that each box would face straight toward the so-called {\it off-point} of the subunit. This is a 
point on the symmetry axis of the subunit at which the symmetry axes of the four boxes intersect, and that is about 3088~mm away from the front face of the subunit. This value is exactly the 
same in both the mechanical and the simulation design, and can be deduced from the distances and angles given in Fig.~\ref{subunit-dimensions}, since 
$[24.37+(0.36/2)+0.24+(0.6/2)]~{\rm mm}/\tan[(180^{\circ}-179.069^{\circ})/2]=3088$~mm. The considered spaces between the crystals in a box and between the box and the alveole 
walls are reserved for the wrapping material of about 0.07~mm thickness as well as a tolerance space to be filled with air.

\begin{figure}
\begin{center}
\resizebox{0.48\textwidth}{!}{\includegraphics{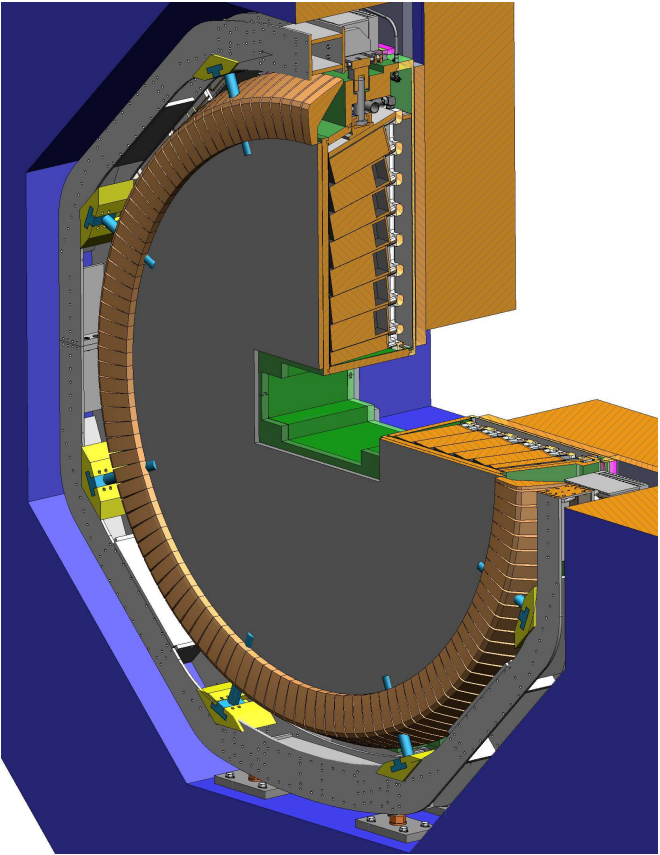}}
\caption{A view of the complete mechanical design of the FwEndCap with thermal insulation cover and holding structure.
The Disc DIRC Cherenkov detector is placed in front of the crystals in the same frame which is inserted in the solenoid magnet of the target spectrometer.}
\label{completeFwEndCap}
\end{center}
\end{figure}
\par For the ease of construction, the mechanical design of the FwEndCap with a nearly circular periphery was constrained to contain only full- or half-subunit packages of crystals. It will 
contain 214 complete subunits and 54 half-subunits in total, which sums up to the total number of 964 (3856) boxes (crystals) in the FwEndCap. Figure~\ref{subunit-dimensions} quantifies various 
gaps between the crystals and thicknesses of alveole walls in different parts of a subunit. Figure~\ref{completeFwEndCap} shows a view of the complete mechanical design of the 
FwEndCap with thermal insulation cover and other details. Although the FwEndCap is built with tapered crystals, the subunits are arranged in a quasi-planar geometry perpendicular to 
the beam axis, in order to ensure a compact setup inside the solenoid.
Consequently, the subunits at larger angles are pulled downstream such that the front tip of at least one crystal stays on one plane and the front tips of all other crystals stay 
behind this plane, defined as the {\it front plane} (see Fig.~\ref{Calorimeter}). Every full- or half-subunit is to be fixed onto a $30$~mm thick aluminium backplane by pre-angled interface pieces 
(see Fig.~\ref{subunit-explosion}) to keep the exact position and orientation. The optimized orientation of the aluminium interfaces and, correspondingly, subunits and half-subunits are 
calculated with UGS NX 5~\cite{Lindemulder}. These mechanical-design values have been taken to implement the geometry of the FwEndCap into the simulation code.

\subsection{The FwEndCap implementation in PandaROOT}
\label{FwEndCapSimul_label}
\begin{figure}
\resizebox{0.48\textwidth}{!}{\includegraphics{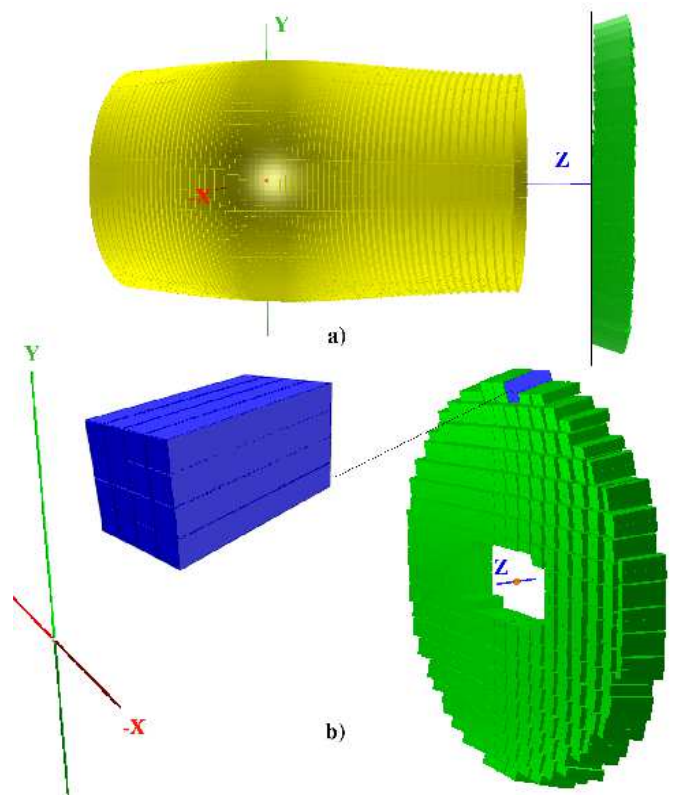}}
\caption{a) A side view of the Barrel and FwEndCap of the $\overline{\rm{P}}$ANDA EMC based on the PandaROOT simulation package. The position of the front plane is shown as a black vertical line. 
b) Geometrical layout of the nearly planar arrangement of the FwEndCap; each block represents a half- or full-subunit; one subunit is shown in an expanded view. This comprises crystals numbered as 
row$=34-37$ and column$=3-6$ in the simulation code.}
\label{Calorimeter}
\end{figure}

\begin{figure}
\begin{center}
\resizebox{0.48\textwidth}{!}{\includegraphics{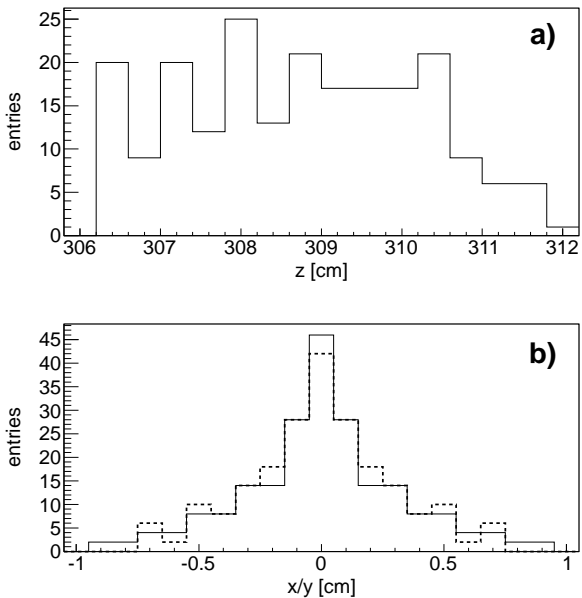}}
\caption{The distribution of the points of closest approach of the symmetry axes of all full-subunits of the FwEndCap to the beam axis; 
a) The distance distribution of the points of closest approach in beam direction with respect to the front plane of the FwEndCap (off-point value). The mean value of this distribution is 3088~mm, as designed;
b) The distribution of the points of closest approach in $x$- (solid line) and $y$-direction (dashed line).}
\label{offpoint_xyz}
\end{center}
\end{figure}

Figure~\ref{Calorimeter} shows the Barrel and FwEndCap of the EMC as drawn within the PandaROOT framework. The maximum polar angle coverage of the FwEndCap is calculated to be 
$24^{\circ}$. This would contain 72 columns and 74 rows of crystals of the FwEndCap. Within the simulations, the crystal column and row numbers range, respectively, from $-36$ to $36$ and $-37$ 
to $37$, excluding $0$.
Figure~\ref{offpoint_xyz} shows the distribution of the points of closest approach of the symmetry axes of the subunits to the beam axis.
The distance distribution of the points of closest approach with respect to the front plane of the FwEndCap is shown in Fig.~\ref{offpoint_xyz} a). 
We call this distance the {\it off-point value} of the subunit, which 
is slightly different for various subunits because of the planar construction. In principle, we expect 57 different off-point values due to the symmetry of the FwEndCap design. The target 
point has a distance of 2039~mm to the front plane of the FwEndCap, while the mean value of the off-point-value histogram indicates a distance of about 3088~mm from the front plane of the 
FwEndCap. This value is the same as the distance of the off-point of the subunit to its front face. Figure~\ref{offpoint_xyz} b) shows the x/y distributions of the points of closest approach.
\par Orienting the crystals towards a confined region (the off-point region) upstream of the target point prevents photons from passing through the gaps between neighboring 
crystals and subunits. The distance between the target point and the off-point region is optimized such that photons should cross a gap between crystals within a distance 
of about 1/3 of the crystal length. This condition has been verified in the simulations and enhances the efficiency of photon detection by the FwEndCap. 

\section{The FwEndCap simulation results}
\label{Simulations_label}
The simulation studies for the $\overline{\rm{P}}$ANDA EMC are focused on the expected performance of the planned EMC with respect to the energy and spatial resolution of the 
reconstructed photons, on the capability of electron/hadron separation, and also on the feasibility of the foreseen physics program of $\overline{\rm{P}}$ANDA~\cite{PandaTechDesRep}.
The simulation package contains event generators with proper decay models for all particles and resonances involved in the individual physics channels~\cite{EvtGen}, particle tracking through the 
complete $\overline{\rm{P}}$ANDA detector, a {\it digitization} procedure which models the signals and the signal processing in the front-end electronics of the individual detectors as well as the 
reconstruction of charged and neutral particles. 
In this Section, the simulations were performed with photons emitted isotropically from the target point, unless noted otherwise.

\subsection{Digitization procedure}
\label{FwEndCap_digitization_label}
The digitization of the EMC has been implemented with realistic properties of PWO crystals at the operational temperature of 
$-25^{\circ}$C~\cite{PandaTechDesRep}. A photon entering one crystal of the EMC develops an electromagnetic shower which extends over several crystals. 
We will refer to a crystal as a {\it hit} if its signal at the output of the simulation has a non-zero value.
Correspondingly, we refer to such a crystal as a {\it digi} if its signal would pass the digitization procedure (the conversion into electronic signal) in which a minimum {\it detection 
threshold} of the order of $2-3$~MeV is required. 
The collection of all the neighboring crystals classified as {\it digi} is called a {\it cluster}, where it is possible for a cluster to contain only a single crystal.
\par Clusters have to be reconstructed with good energy, position, and time resolutions over a wide dynamic range starting from about 10 MeV up to 12~GeV, as expected in the 
reactions to be studied with $\overline{\rm{P}}$ANDA. In these simulations, the statistical fluctuations are estimated, assuming a light yield of 500 photons/MeV for the crystals and a quantum 
efficiency of 0.7 and 0.22 for the LAAPD and VPTT photo-sensors, respectively~\cite{PandaTechDesRep}. 
Based on the above-mentioned PWO light yield at the operation temperature, a FwEndCap crystal with the back-face area of $26\times26$~mm$^{2}$ and equipped with a VPTT of 200~mm$^2$ cathode 
area would yield 33 photo-electrons/MeV. As this is the worst-case operating condition because of the low quantum efficiency, we adopt this case for the further performance estimates. 
The signal is convolved with a Gaussian distribution with a standard deviation of $\sigma = 1$~MeV to account for the electronic noise of an individual detector element. This noise level is a conservative 
estimate based on measurements with a digital readout scheme and digital data treatment \cite{Kavatsyuk}. It includes the contributions of the LAAPD excess-noise factor folded with the 
electronic noise of the preamplifier and readout electronics based on the Sampling ADC (SADC) technique. 
In the case of VPTT, the noise contribution is dominated by the electronic noise of the preamplifier, yielding about the same performance as the LAAPD readout. 
\par In order to correctly retrieve the crystal position from its row/column number, as indicated by the {\it hit}, the proper mapping of the geometry has been ensured in the simulations. 
A monotonous and approximately linear correlation between crystal row number and $y$-position (or between crystal column number and $x$-position) of the geometrical center of the crystal 
was observed which is expected by design of the crystal arrangement in the FwEndCap. 
Depending on the studied cases, either only the FwEndCap module or the full EMC, including FwEndCap, BwEndCap, and Barrel EMC modules, will be present in the simulated geometry.

\subsection{Cluster reconstruction}
\label{FwEndCap_cluster_label}
A photon impinging on the EMC 
could initiate several (disconnected) clusters. There could be various reasons for the occurrence of disconnected clusters, so-called {\it split-offs}: migration of a secondary particle, created in 
one cluster, to another part of the EMC, but not through the crystal material; statistical 
fluctuations leading to a probability for particles to pass through the crystals without interacting with the bulk of crystals on their way. Also, due to fluctuations in the hadronic and 
electromagnetic shower-energy distribution, too many clusters may be reconstructed in the EMC, which are not associated with separate primary particles. The shower-shape analysis allows to 
discriminate such split-off clusters from the photon clusters. One can effectively suppress the split-off clusters by choosing a proper cluster threshold for the physics channel of 
interest. Although this will be our approach in this paper, there is a need for a more refined algorithm exploiting the threshold dependence.
\par In general, one can assume two different kinds of split-off clusters: the electromagnetic split-off is a cluster of crystals (usually one) that is located near the primary photon cluster 
and produced by a shower product (e.g. photon) which interacts in a crystal not connected directly to the primary cluster of crystals.
The hadronic split-off results in a cluster of crystals (usually very few) produced by some secondary particle (e.g. neutrons) due to 
the interaction of a primary charged particle ($\pi$, $K$) somewhere in the detector. As a result, this type of split-off looks like a photon and perturbs the analysis of events. The 
characteristics of hadronic split-offs were studied by the Crystal Barrel collaboration \cite{Burchell}. From these analyses of the hadronic split-offs one can conclude the following:
\begin{itemize}
\item They pose a serious problem, since one can expect a 50\% chance of split-off per charged hadron;
\item Below 20 MeV, hadronic split-offs are very prolific;
\item About half of the split-offs are not located near a primary charged particle track;
\item There seem to be no good criteria to flag split-offs.
\end{itemize}

\begin{figure}
\begin{center}
\resizebox{0.48\textwidth}{!}{\includegraphics{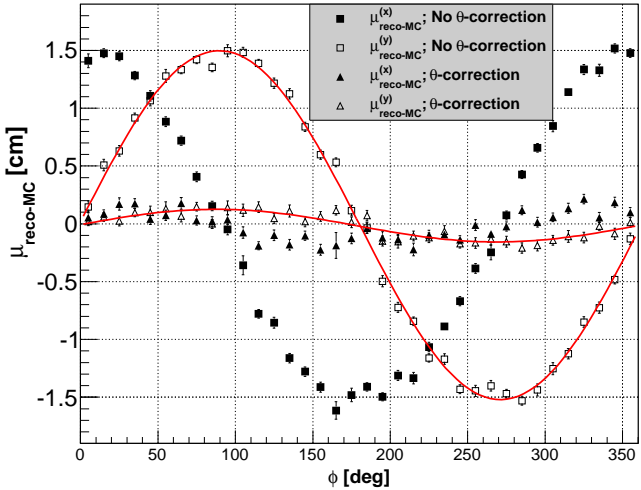}}
\caption{The mean value ($\mu_{reco-MC}$) of the Gaussian fit to the histogram $y_{cl}-y_{MC}$ (or $x_{cl}-x_{MC}$) for the cluster with highest energy deposition per event which satisfies 
the threshold condition of 30~MeV. $A$ and $B$ in Eq.~\ref{posOffset} are assumed here to be 3.8 and 0, respectively, and the isotropically-generated photons of 100~MeV scan 
over the FwEndCap. The error bars are the fit errors of the mean values. The curves are sinusoidal fits to the data; the data are shown for cases with or without using
the lookup table for the correction of the $\theta$-angle.}
\label{clusterThetaCorrection}
\end{center}
\end{figure}
\begin{figure}
\begin{center}
\resizebox{0.48\textwidth}{!}{\includegraphics{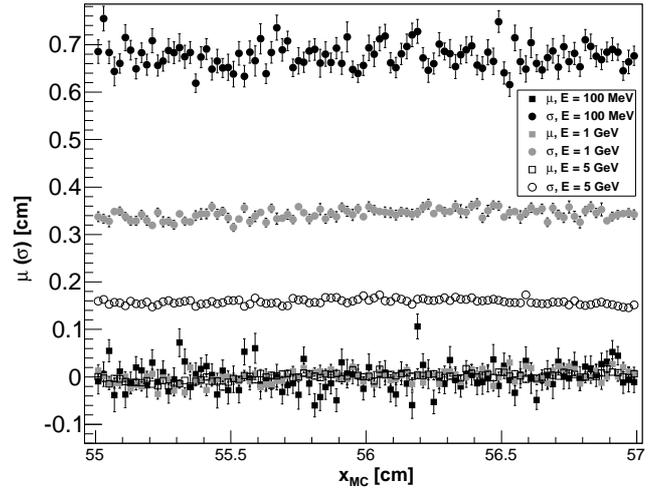}}
\caption{The mean ($\mu$) and standard deviation ($\sigma$) of the Gaussian fit to the histogram $y_{cl}-y_{MC}$ for the cluster with highest energy deposition per event which satisfies the 
threshold condition of 30~MeV. Here, the isotropically-generated photons of 100~MeV, 1~GeV, and 5~GeV scan over a region covering the gap between two neighboring subunits ($x\approx 56$~cm).}
\label{subunitGapSurvey}
\end{center}
\end{figure}
\subsubsection{Cluster position}
\label{FwEndCap_clusterposition_label}
\par In this analysis, the FwEndCap is assumed to be the only module present in the simulated geometry. 
The event analysis is performed only for clusters with highest energy deposition that satisfy a threshold condition of 30~MeV. 
In Section~\ref{FwEndCap_clustermultiplicity_label} it will be shown that a 30~MeV threshold results in a lower deficit in the number of registered low-energy photons as 
compared to 1~GeV photons. This threshold value limits the maximum number of registered clusters for 1~GeV photons to 2 per event. 
Events with two registered clusters comprise only 0.2\% of the total number of events for which at least one cluster satisfies the threshold. 
The $x$-/$y$-position of photons hitting the FwEndCap is reconstructed through projecting the reconstructed position of the cluster ${\overrightarrow{r}}_{cl}=(x_{cl},y_{cl},z_{cl})$ 
along the axis and on the front face of the crystal with the highest deposited energy in the cluster.
$\overrightarrow{r}_{cl}$ is defined to have a linear relation with respect to all the {\it digi} positions ($\overrightarrow{r}_{digi}$). 
Simulation studies of 1~GeV photons show a mean shower depth of 62~mm, called the {\it digi} depth-position. 
The {\it digi} depth-position is defined as the distance of $\overrightarrow{r}_{digi}$ (the point located on the crystal axis) to the center of the front face of the crystal.
This would then give better reconstruction of the impact point of photons than the geometrical center of the crystal ({\it hit} position).
In Section~\ref{FwEndCap_correction_label}, we will use an optimized lookup table to correctly reconstruct the position of the clusters.
The obtained value of the {\it digi} depth-position was used as the starting point in the simulations to build this lookup table, which will then take care of
the deviations for various photon energies. The look up table will eventually absorb irregularities like the shower leakages as well as the assumed 
crystal properties like non-uniformity in the light collection, the {\it digi} threshold, and the {\it digi} depth-position.
The simulation analysis shows that the position reconstruction
in the FwEndCap gets worse when one uses a linear energy weighting instead of a logarithmic one. This is shown to be more crucial for 1~GeV as compared to 100~MeV photons.
The position of the cluster $\overrightarrow{r}_{cl}$ is thus obtained through a logarithmic rather than a linear energy weighting formula.
This ansatz is motivated by the fact that the shower profile has an exponentional shape.
Hence, every individual crystal with $E_{digi}$ energy deposition belonging to a cluster with an energy content of $E_{cl}$ would contribute a positive weight of $W_{digi}$ in the position 
reconstruction as $W_{digi} = W_0+\ln(E_{digi}/E_{cl})$, where $W_0$ would control the smallest fractional energy that a crystal can have and still contribute to the position measurement. 
\begin{figure*}
\begin{center}
\resizebox{0.9\textwidth}{!}{\includegraphics{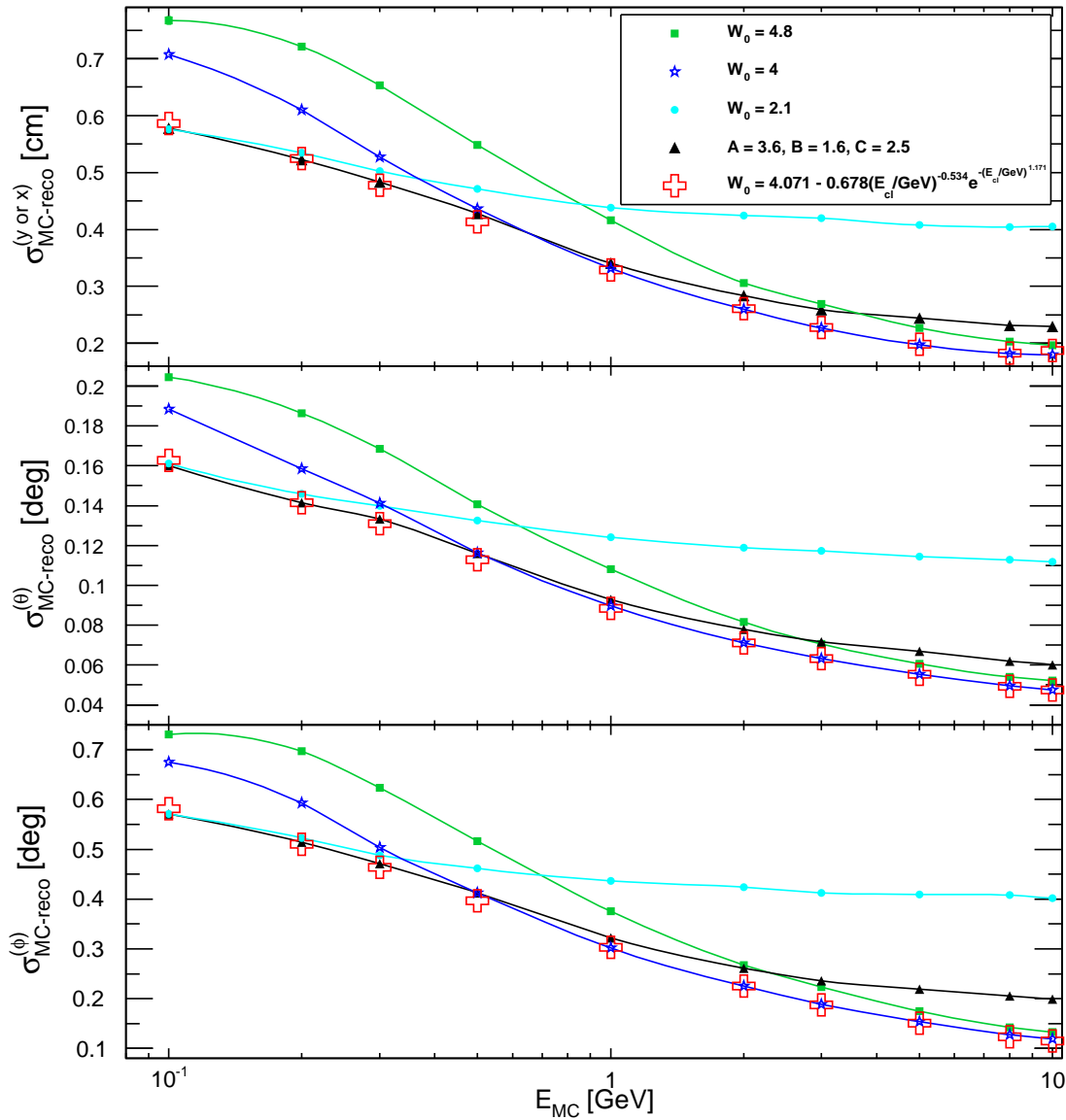}}
\caption{Simulation results for the $\sigma=$~FWHM/2.35 of the difference between the reconstructed and Monte Carlo values of the $x$-/$y$-position (top panel) and the 
$\theta$-/$\phi$-angle (bottom panels) of the registered clusters in the FwEndCap as a function of the photon energy. Data are provided for various values of $W_0$ as well as one 
set of the parameters $A$, $B$, and $C$ that are related to a logarithmic weighting for position reconstruction, as introduced in Eq.~\ref{posOffset}.
The hollow pluses show the results taking Eq.~\ref{NEWposOffset} as the functional form for $W_0$.
The error bars show the statistical uncertainty and are barely visible. The photons are generated within $\theta=15^{\circ}-18^{\circ}$ over the FwEndCap. 
A smooth line is drawn through each data set to guide the eye.}
\label{thetaPhiRes_photons}
\end{center}
\end{figure*}
As $W_0$ decreases, the modules with the highest energy are weighted more heavily, while modules having an energy fraction below $\exp(-W_0)$ are excluded entirely. 
Therefore, if $W_0$ is too small, only a few modules dominate the position 
calculation and the results become sensitive to the position of incidence and the distribution of the position difference becomes non-Gaussian \cite{Awes}.
For instance, for $E_{cl}=30$~MeV (100~MeV), only a $W_0\geq 2.3~(3.5)$ guarantees that contributions of 
crystals with $E_{digi}\geq 3$~MeV will be taken into account in the position reconstruction of the cluster. A $W_0=2.3~(3.5)$ means that only crystals in the cluster containing more than 10\% (3\%) of 
the cluster energy will contribute to the position measurement. In the PandaROOT package, the functional form of $W_0$ has been taken to be the same as was used by the 
BaBar experiment~\cite{BaBarExp}:
\begin{equation}
W_0= A - B\cdot \exp(-C\cdot E_{cl}).
\label{posOffset}
\end{equation}
If we would have more than one {\it digi} (crystal) in the cluster, the cluster position would be 
\begin{equation}
\overrightarrow{r}_{cl} = \frac{\displaystyle\sum\limits_{digi}{W_{digi}\cdot~\overrightarrow{r}_{digi}}}{\displaystyle\sum\limits_{digi}{W_{digi}}}.
\label{clustpos}
\end{equation}
\par In reconstructing the position of the photons, there is a $\phi$-dependence observed for the mean values $\mu_{reco-MC}$ of the deviation between reconstructed and MC-true positions, 
expressed by Gaussian fits to the histograms $y_{cl}-y_{MC}$ and $x_{cl}-x_{MC}$. This $\phi$-dependence of sinusoidal shape (see Fig.~\ref{clusterThetaCorrection})
is a geometrical effect because the off-point and the target point do not coincide.
Since the off-point is located about 1~m upstream of the target point, most photons generated at $\phi\approx 0^{\circ}$, cause a horizontal rather than a vertical shift of the shower position. This effect
would correspond to a minimum (maximum) expected value for $|y_{cl}-y_{MC}|$ at $\phi\approx 0^{\circ}$ ($\phi\approx 90^{\circ}$). Similarly, based on symmetry arguments, we should observe 
a minimum (maximum) expected value for $|x_{cl}-x_{MC}|$ at $\phi\approx 90^{\circ}$ ($\phi\approx 180^{\circ}$), which ensures the sinusoidal behavior.
\begin{figure}
\begin{center}
\resizebox{0.48\textwidth}{!}{\includegraphics{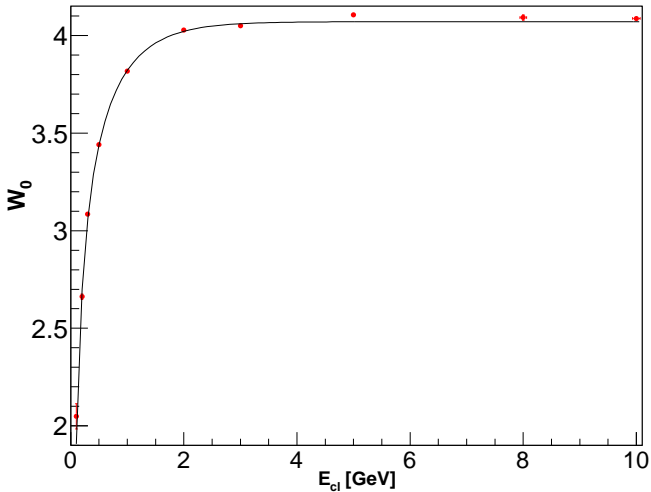}}
\caption{Optimized $W_0$ derived for the FwEndCap as a function of cluster energy. The curve is a fit with the functional form given in Eq.~\ref{NEWposOffset}. The horizontal error bars 
that are barely visable indicate the $\sigma=$~FWHM/2.35 of the difference between the photon energy and the reconstructed energy of the cluster with highest deposited energy. The value 
of the optimized $W_0$ was taken as the one corresponding to the minimum resolution by fitting the resolution as a function of $W_0$. The error size of $W_0$ was calculated from the fit, 
taking into account the $\chi^{2}$ of the fit for each energy. The reduced $\chi^{2}$ of the fit in this Figure is $4.54$.}
\label{W0_label}
\end{center}
\end{figure}
\par Due to the fact that the off-point and target point do not coincide, a lookup table has been exploited for 
the correction of the cluster $\theta$-angle that is derived from the Monte Carlo studies. The reconstruction of the cluster position shows clear improvement when the cluster $\theta$ angle is 
corrected by the lookup table. This lookup table will be discussed in Section~\ref{FwEndCap_correction_label} along with the lookup table for the energy correction. For the rest of the 
discussion we apply this correction.
The residual deviation of the corrected mean difference $\mu_{reco-MC}$ in Fig.~\ref{clusterThetaCorrection} (100~MeV photons) from the zero level amounts to a small uncertainty in
the position resolutions that will be presented below. This is calculated to be about 0.1~mm, 0.13~mm, and 0.2~mm for photon energies of 100~MeV, 1~GeV, and 5~GeV, respectively.
\par For various photon energies, Fig.~\ref{subunitGapSurvey} shows the mean ($\mu$) and standard deviation ($\sigma$) of the Gaussian fit to the histogram $y_{cl}-y_{MC}$ for 
the cluster with highest energy deposition per event which satisfies the threshold condition of 30~MeV. Photons 
are generated over $15^{\circ}<\theta_{MC}<16^{\circ}$ and $7^{\circ}<\phi_{MC}<10^{\circ}$. This solid-angle range allows for scanning a region covering the gap between two neighboring 
subunits. This is the gap between the crystal columns \#22 and \#23, which is located at $x=56$~cm. 
$x_{MC}$ and $y_{MC}$ are the coordinates of the entrance point of the photon to the front face of either of the two neighboring crystals that is under scanning. 
The scanning steps in the $x$-direction are taken to be 0.2~mm, whereas the thickness of 
the alveole between the two subunits is 0.36~mm. The continuity of $x_{MC}$ dependence of the resolution results ($\sigma$ in Fig.~\ref{subunitGapSurvey}) demonstrates that 
there will not be local inhomogeneities that could introduce fluctuations in the position resolutions for individual photons. 
\begin{figure*}
\begin{center}
\resizebox{0.9\textwidth}{!}{\includegraphics{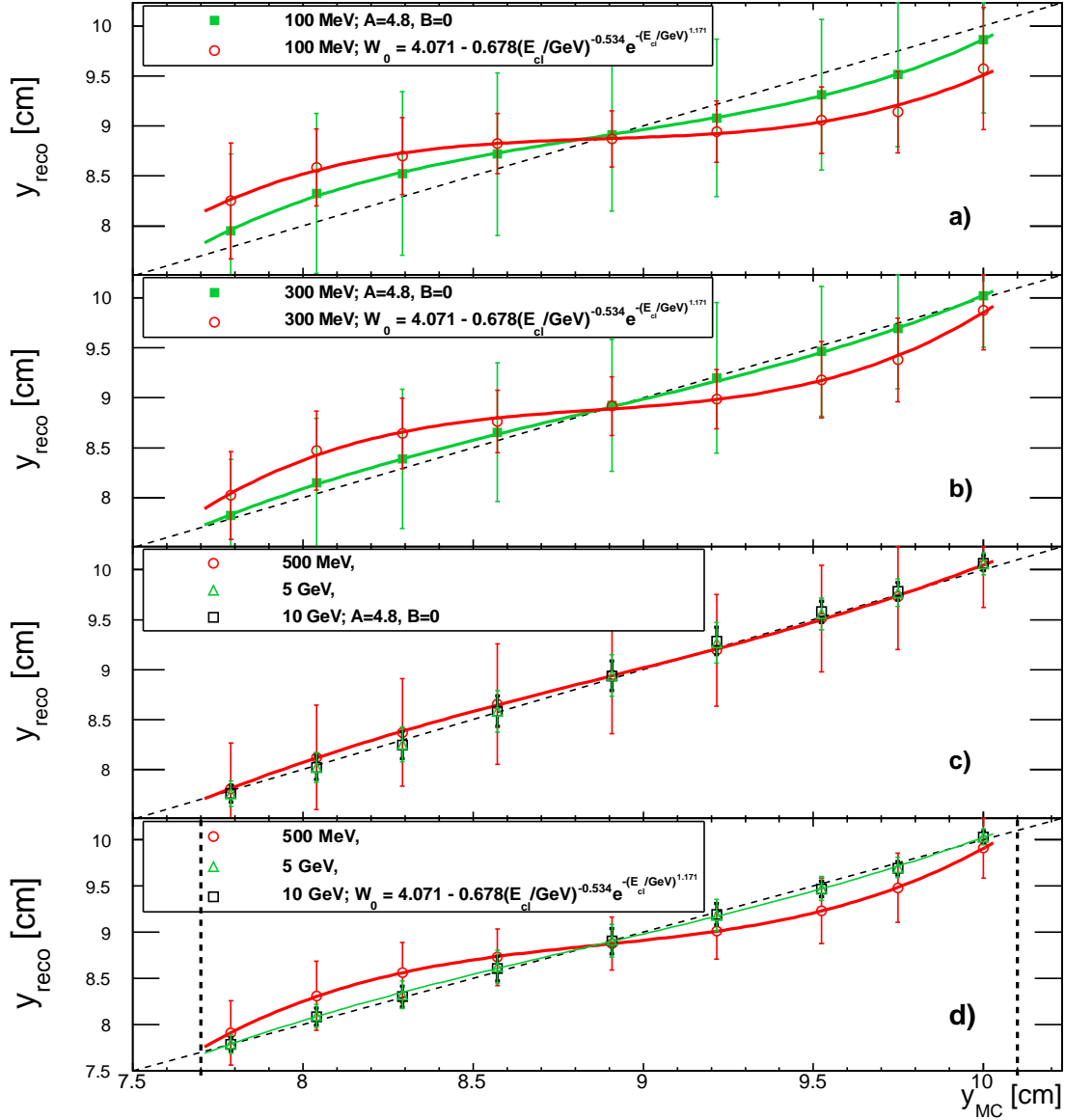}}
\caption{The reconstructed mean values versus Monte Carlo values of the photon position for nine equidistant hitting points along the $y$-axis at the mid-$x$ on the face of the crystal located 
at [row=4,column=25]. The vertical dashed lines mark the left and right edges of the front face of the crystal. The error bars are the standard deviation of the reconstructed $y$-position. The 
data are for five photon energies, one set of parameters in Eq.~\ref{posOffset} (panels a-c) and the parametrization in Eq.~\ref{NEWposOffset} (panels a, b, and d). The curves are fits to the data points.}
\label{S_curve}
\end{center}
\end{figure*}
\par In the PandaROOT simulations of the FwEndCap, we optimized the parameter $A$ in Eq.~\ref{posOffset} for a few values of $B$ and $C$, and for the energy range 
from 100~MeV to 10~GeV. The sensitivity of the simulations was checked by allowing $A$ to change from 2 to 5 in steps of 0.1 as well as from 6 to 10 in steps of 1.
For instance, for $C=2.5$ and $B=1.6$, the sensitivity was checked resulting in the optimized value for $A$ of 3.6. 
Figure~\ref{thetaPhiRes_photons} compares the corresponding resolution results with the ones derived assuming energy-independent $W_0$ values. 
The resolution results are limited to a central region of the FwEndCap, where there is little chance for shower leakage from the boundaries of the FwEndCap.
Thus, we observe the intrinsic resolution while distorting effects have to be taken into account lateron.
One can see from Eq.~\ref{posOffset} that, as the energy increases, the exponential term diminishes the contribution of the parameter $B$. 
In optimizing these parameters for various modules of the EMC, one notices that the influence of $B$ and $C$ would become appreciable for the low energy clusters. In principle, one 
needs to find the optimized values of $B$ and $C$ in conjunction with $A$ if one employs the parameterization of Eq.~\ref{posOffset}. Alternatively, one can assume an 
energy-dependent $W_0(E_{cl})$ and try to find its functional form by allowing $W_0$ to vary within a reasonable range.
Figure~\ref{W0_label} shows the optimized $W_0$ derived for the FwEndCap as a function of the cluster energy. For each photon energy, $W_0$ was changed between 2 and 5 in steps 
of 0.05 and the optimized value of $W_0$ was chosen to be the one that resulted in the best position resolution of the cluster with highest deposited energy.
Using this $W_0(E_{cl})$ function for the FwEndCap in the absence of any potential non-uniformity in the light yield, the resolution results are presented in Fig.~\ref{thetaPhiRes_photons} (hollow pluses).
These results suggest the following parametrization, obtained by a fit with $\chi^{2}_{R}=4.54$, for the logarithmic energy weighting to be used for the FwEndCap instead of Eq.~\ref{posOffset}:
\begin{equation}
W_0=4.071-0.678\times \left(\frac{E_{cl}}{{\rm GeV}}\right)^{-0.534}\cdot \exp\left(-\left(\frac{E_{cl}}{{\rm GeV}}\right)^{1.171}\right).
\label{NEWposOffset}
\end{equation}
\par The $x$-/$y$-position uncertainties in Fig.~\ref{thetaPhiRes_photons} appear to be small compared to the width (24.37~mm) of the front face of the crystals. The reason for a worse 
resolution in the $\phi$ reconstruction than in the $\theta$ reconstruction is an average lower number of crystals per unit $\phi$-angle than per unit $\theta$-angle. These 
angular resolutions should be compared with the angular coverage of a single crystal seen from the target point, which varies, for different crystals of the FwEndCap, within the 
ranges $0.6^{\circ}<\Delta\theta<0.7^{\circ}$ and $1^{\circ}<\Delta\phi<7^{\circ}$ along the $\theta$- and $\phi$-direction, respectively.
\par Figure~\ref{S_curve} shows the reconstructed versus Monte Carlo mean values of the photon position for nine equidistant coordinates of impact along the $y$-axis on the face of the crystal. 
The results are shown for five photon energies, to be representative of typical low- and high-energy photons in the $\overline{\rm{P}}$ANDA experiment, and one set of parameter values 
in Eq.~\ref{posOffset} ($A=4.8$ and $B=0$). 
The FwEndCap analyses showed that the closest mean values of the reconstructed $x$-$/y$-position to the diagonal line in Fig.~\ref{S_curve} would correspond to $A=4.8$ and $B=0$. This is checked 
to be the case for various photon energies. As the photon energy increases, the curved function fitted to the mean values tends to become 
straight along the diagonal line. However, the error bars are considerably larger than those derived from Eq.~\ref{NEWposOffset} (compare, for instance, Fig.~\ref{S_curve} a) and b)), in accordance with 
the corresponding resolutions shown in Fig.~\ref{thetaPhiRes_photons}. 
For high-energy photons of 10~GeV, as opposed to an energy of 5~GeV, the smaller size of the error bars in Fig.~\ref{S_curve} ensures higher resolutions in Fig.~\ref{thetaPhiRes_photons}. 
\begin{figure*}
\begin{center}
\resizebox{0.9\textwidth}{!}{\includegraphics{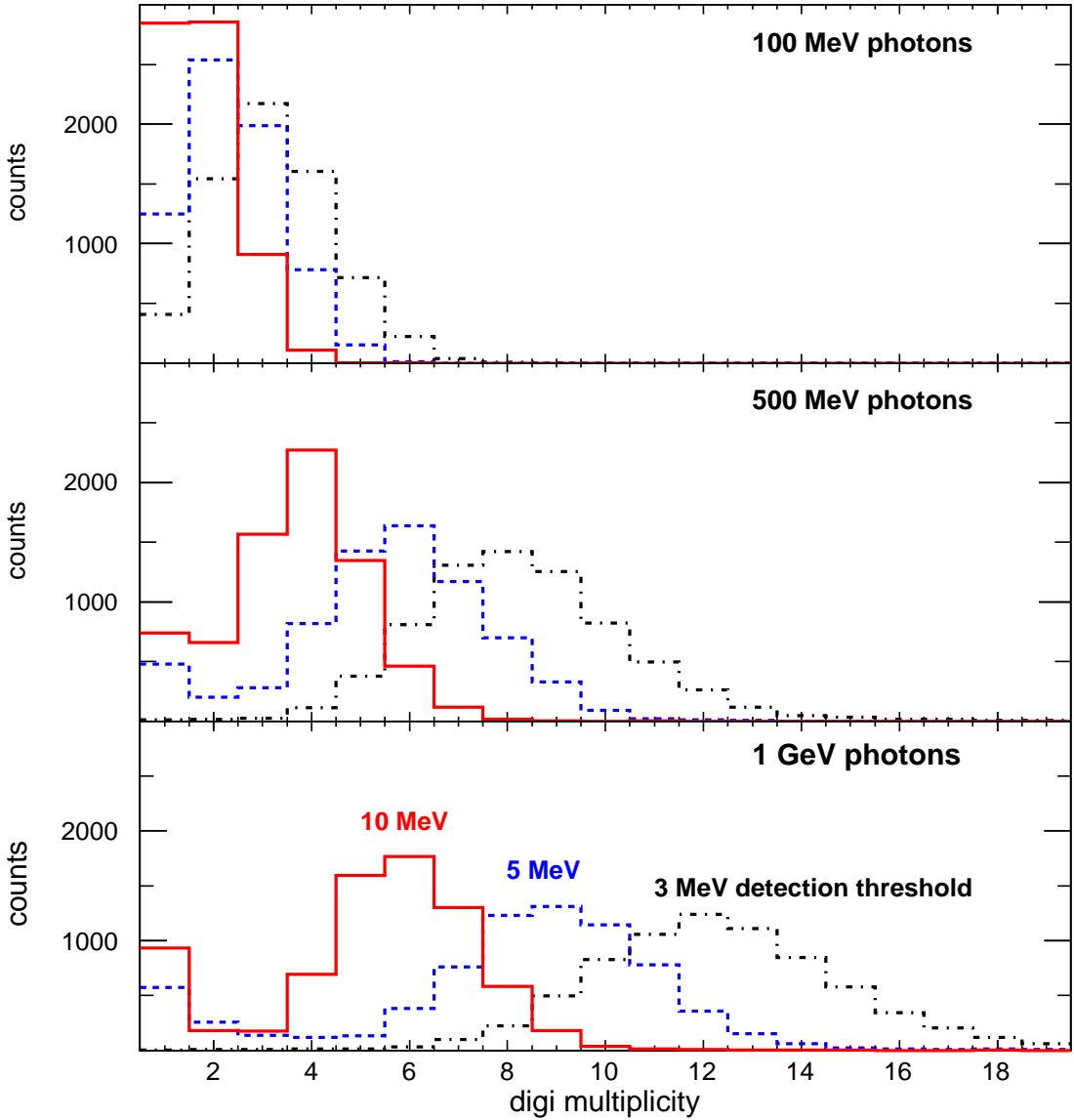}}
\caption{Sensitivity of the {\it digi} multiplicity distribution in the FwEndCap module to various detection thresholds of 3~MeV (dash-dotted), 5~MeV (dashed blue), and 
10~MeV (solid red histograms) for 100~MeV (top panel), 500~MeV (middle panel), and 1~GeV (bottom panel) photons emitted into the full EMC.
}
\label{multiplicitySensitivity}
\end{center}
\end{figure*}
\subsubsection{Cluster multiplicity}
\label{FwEndCap_clustermultiplicity_label}
In this analysis, the FwEndCap module is embedded in the full EMC geometry. Figure~\ref{multiplicitySensitivity} shows the sensitivity of the {\it digi} multiplicity distribution to various 
detection thresholds for photons of 100~MeV, 500~MeV, and 1~GeV energy.
\par
The mean values $N$ of the {\it digi} multiplicity distributions are summarized in Table~\ref{averageDigis}.
\begin{table}
\caption{The mean $N$ of the {\it digi} multiplicity distributions shown in Fig.~\ref{multiplicitySensitivity} for various detection thresholds and photon energies. In addition to 
Fig.~\ref{multiplicitySensitivity} the mean values for the 0 MeV detection thresholds are also shown here.
}
\begin{center}
\begin{tabular}{ | c | c | c | c | c | } \hline
   &\multicolumn{4}{c|}{$N$} 			    \\
   photon energy [MeV]&\multicolumn{4}{c|}{for detection threshold [MeV]} \\ 
		& 0 	&  3 	& 5   & 10     \\ \hline
    100         &  6.1	&  3.2 	& 2.4 & 1.7    \\
    500         & 16.1  &  8.3 	& 5.6 & 3.7    \\
   1000         & 23.2  & 12.6	& 8.1 & 5.3    \\ \hline
\end{tabular}
\label{averageDigis}
\end{center}
\end{table}
In the case of a $0$~MeV detection threshold, the mean {\it digi} multiplicity increases significantly. 
The relatively high number of events with one {\it digi} per event, as compared to two or three {\it digi}s per 
event, for the multiplicity patterns of 5 and 10~MeV detection thresholds in the lower two panels of Fig.~\ref{multiplicitySensitivity}, reflects the fact that a considerable number of 
the FwEndCap crystals close to the circumference are screened against direct impact of photons by the forward edge of the Barrel EMC (see Fig.~\ref{Calorimeter}). This overlap is intended 
in order to guarantee a complete coverage for photon detection.
\begin{figure*}
\begin{center}
\resizebox{0.9\textwidth}{!}{\includegraphics{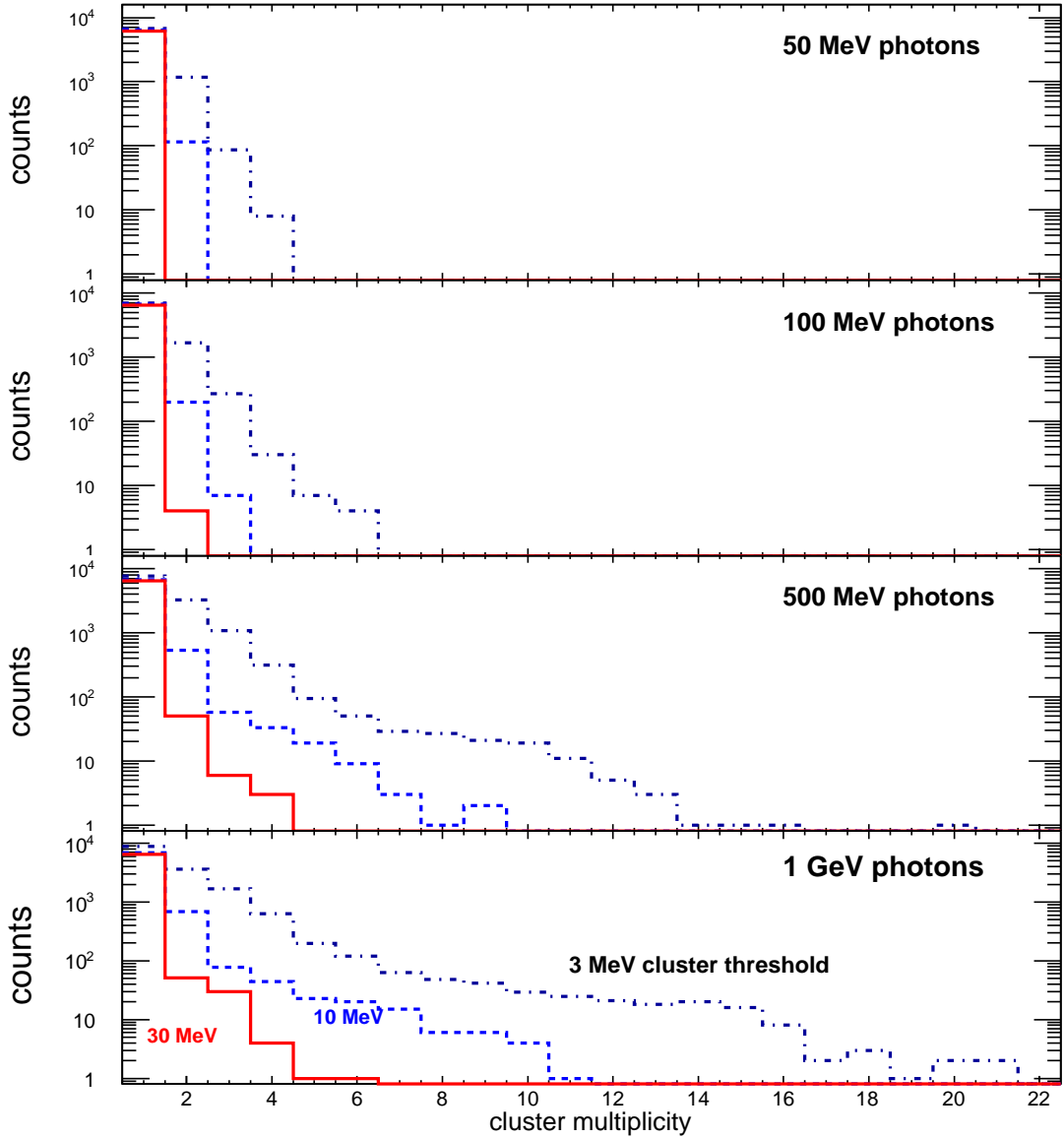}}
\caption{Cluster multiplicity distributions in the FwEndCap for various cluster thresholds of 3~MeV (dash-dotted black), 10~MeV (dashed blue), and 30~MeV (solid red histograms) for photons 
of 50~MeV, 100~MeV, 500~MeV, and 1~GeV energy. Here, the digitization detection threshold is 3~MeV.}
\label{clusterMultiplicity}
\end{center}
\end{figure*}
High-energy photon showers that partially leave the forward edge of the Barrel EMC would mostly cause one-{\it digi} clusters in the FwEndCap for detection thresholds above 3~MeV. On the other 
hand, photons that directly hit the FwEndCap would cause a Poissonian {\it digi}-multiplicity distribution. The pattern that we observe in the lower panel of Fig.~\ref{multiplicitySensitivity} 
for the 10~MeV detection threshold is a superposition of the Poissonian and the one-{\it digi} distributions.
\par Figure~\ref{clusterMultiplicity} shows the multiplicity of clusters for the FwEndCap module, assuming various cluster thresholds for 50~MeV, 100~MeV, 500~MeV, and 1~GeV photons 
emitted into the full EMC geometry. As mentioned in Section~\ref{FwEndCap_clusterposition_label}, if the FwEndCap is the only module present in the simulated geometry, we would expect 
for 1~GeV photons and a 30~MeV detection threshold a maximum of two clusters per event, where about 0.2\% of the events contain 2 clusters. 
In the corresponding cluster multiplicity distribution of Fig.~\ref{clusterMultiplicity} for the full geometry, the percentage of events with 2 clusters is 0.8\% due to the effect 
of the forward edge of the Barrel increasing the cluster multiplicity in the FwEndCap. We have checked for the optimum value of the cluster threshold energy (as far as the 
efficiency is concerned), by comparing the results for 50~MeV, as representative for low-energy photons, with the ones for 1~GeV photons. For various cluster thresholds, 
we studied the ratio of the number of events with cluster multiplicity of 1 for 50~MeV photons to the number of events with cluster multiplicity of 1 for 1~GeV photons.
Table~\ref{deficitClusterNo} summarizes the obtained simulation results for the FwEndCap as part of the full EMC module.
It shows that, assuming a 30~MeV rather than other cluster thresholds, one would expect a lower deficit in the number of the registered low-energy (say, 50~MeV) photons as compared 
to 1~GeV photons.
\begin{table}
\caption{Simulation results of the FwEndCap, as part of the full EMC module, showing the decrease in the percentage of events with cluster multiplicity of 1, when 
comparing 50~MeV with 1~GeV photons.}
\begin{center}
\begin{tabular}{ | c | c | c | c | c | c | } \hline
  cluster threshold [MeV]		& 10 	&  20 	& 30  & 35   & 40  \\ \hline
  Registered event deficit [\%]         & 7     &  4.7	& 4.5 & 6.1  & 12.3  \\ \hline
\end{tabular}
\label{deficitClusterNo}
\end{center}
\end{table}

\subsection{Efficiency and energy resolution for photons}
In this Subsection, the simulated geometry contains only the FwEndCap module in order to evaluate its performance in photon detection without effects of neighbouring detector components.
\subsubsection{Full-energy efficiency}
Since electromagnetic showers could leak 
out of the FwEndCap, thereby losing part of the energy of the impinging photons, we performed a number of simulations to obtain the average collected energy for various photon energies. 
\begin{figure}
\begin{center}
\resizebox{0.48\textwidth}{!}{\includegraphics{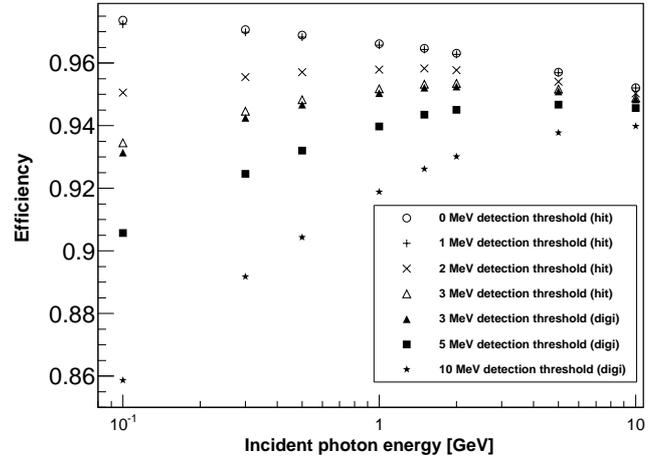}}
\caption{The full-energy efficiency of the sum of energy depositions in the FwEndCap as a function of the thrown photon energy, for photons emitted isotropically from the target point.
For the analysis based on the {\it hit} information, the results are presented for detection thresholds of 
zero (circles), 1~MeV (plus symbols), 2~MeV (cross symbols), and 3~MeV (hollow triangles). For the analysis based on {\it digi} information, the results are presented for detection 
thresholds of 3~MeV (solid triangles), 5~MeV (squares), and 10~MeV (stars).}
\label{fullEnergyEfficiency}
\end{center}
\end{figure}
\par Figure~\ref{fullEnergyEfficiency} shows the {\it full-energy efficiency}, i.e. the fraction of the full energy registered by those crystals which satisfy the detection threshold. We expect 
a considerable fraction of energy leaking out for hitting points close to the edges of the FwEndCap. This 
explains why we see, on average, 3\% loss in the efficiency even at very low energies for the threshold-free case. In a threshold-free situation (circles), detectors fully register the 
deposited energies and, hence, the average percentage of the total registered energy would decrease with increasing photon energy due to the increase in shower leakage from the 
edges. This decreasing trend persists even when we introduce a 1~MeV threshold for the crystals (plus symbols), but starts to change already as we approach a 2~MeV threshold (cross 
symbols). The reason for this change of trend is that at lower energies, say 0.1~GeV, a higher percentage of incident energy is not registered due to the 
threshold condition. At a higher photon energy, say 0.3~GeV, a lower percentage of the incident energy would be lost due to non-firing crystals. Therefore, for 
thresholds higher than about 2~MeV, the overall curve of the full-energy efficiency rises with increasing photon energy and then drops, as can be seen for various detection thresholds in 
Fig.~\ref{fullEnergyEfficiency}. The reason for the subsequent decrease of the efficiency is that for photons with high enough energy the leakage of energy 
from the edges of the FwEndCap would dominate over the aforementioned threshold effect. The various efficiency curves tend to approach each other, as the photon energy increases, due to 
the dominant impact of the energy leakage in determining the full-energy efficiency. Also, the impact of noise diminishes, as is observed for the {\it hit} and {\it digi} curves for 3~MeV detection 
threshold, which smoothly coincide as the energy increases.
  
\subsubsection{Energy resolution}
\begin{figure}
\begin{center}
\resizebox{0.48\textwidth}{!}{\includegraphics{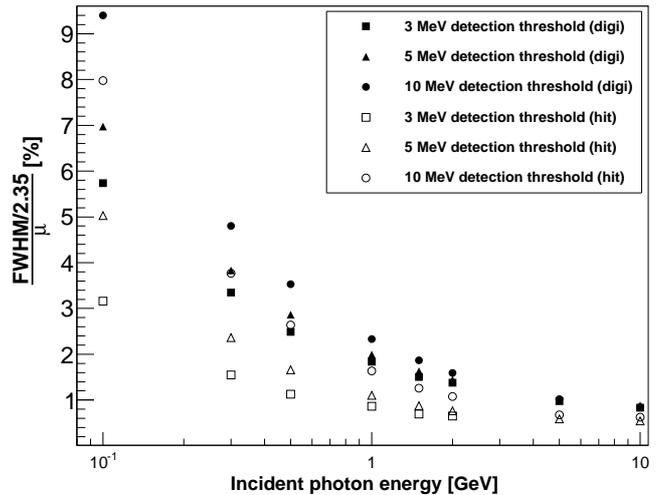}}
\caption{Simulation results for the energy resolution of the FwEndCap as a function of the incident energy of photons.
The simulations are performed for detection thresholds of 3, 5, and 10~MeV at {\it hit} and {\it digi} levels. FWHM and $\mu$ on the y-axis represent the full width at half maximum and the peak 
position, respectively, of the Lorentzian fit function to the total energy deposition registered by the crystals. In these simulations a uniform light collection efficiency at various 
interaction points along the length of the crystals is assumed.}
\label{FwEndCapPhotonResolution_1}
\end{center}
\end{figure}

\begin{figure}
\begin{center}
\resizebox{0.48\textwidth}{!}{\includegraphics{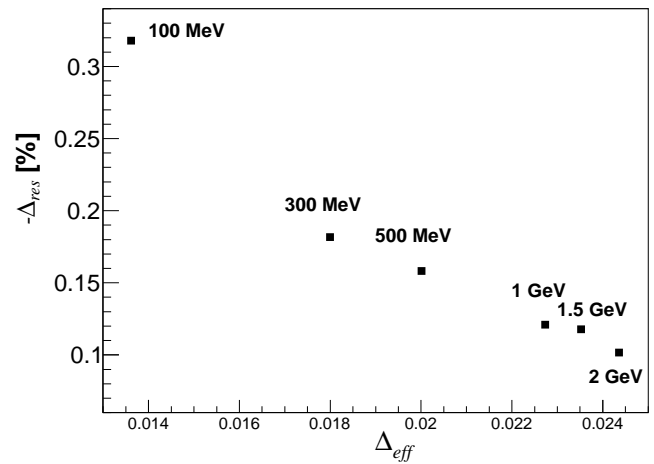}}
\caption{The difference in the energy resolution ($-\Delta_{res}$) as a function of the difference $\Delta_{eff}$ in full-energy efficiency for the two cases of the average over the 
FwEndCap and a well-defined cluster position inside the FwEndCap. The central crystal of the cluster is located at [row=2,column=19]. In these simulations a 3~MeV {\it digi} threshold is 
assumed. The data points correspond to photon energies of 100, 300, 500, 1000, 1500, and 2000~MeV.}
\label{resolution_efficiency}
\end{center}
\end{figure}

Figure~\ref{FwEndCapPhotonResolution_1} shows the simulations for the energy resolution of the FwEndCap calorimeter for various detection thresholds at {\it hit} and {\it digi} levels.
The {\it hit} and {\it digi} concepts were explained in Section~\ref{FwEndCap_digitization_label}, where a {\it digi} was defined as 
a {\it hit} satisfying a certain threshold energy in the presence of noise. Clearly, increasing the detection threshold drastically worsens 
the resolution, especially as we approach the lower photon energies. This reflects the 
pattern observed in Fig.~\ref{fullEnergyEfficiency} and implies that the detection threshold appears to be highly influential on the 
quantities of interest such as the full-energy efficiency as well as the energy resolution. On the other hand, by increasing the photon energy, the effect of the detection threshold (in the order 
of 10~MeV) on estimating the resolution fades away; in the case of 5 and 10~GeV photons, the estimated {\it digi} resolutions practically coincide for the three detection thresholds 3, 5, 
and 10~MeV. It is also worth noting that {\it hit} and {\it digi} resolution curves seem to approach each other, as we increase the photon energy. 
The relative energy resolution is conventionally described by 
$\frac{\sigma}{E} = a \oplus \frac{b}{\sqrt{E}} \oplus \frac{c}{E}$ \cite{PDG} 
with the constant instrumental term, the stochastic term, and the noise term weighted by the parameters $a$, $b$, and $c$, respectively.
The stochastic term includes intrinsic shower fluctuations, photo-electron statistics, dead material at the front of the calorimeter, and sampling fluctuations. The noise term 
includes contributions from the readout electronics noise whose contribution to the standard deviation $\sigma$ does not depend on energy. The constant term includes non-uniformities in signal collection 
and the possible error in the detector calibration. The symbol $\oplus$ represents addition in quadrature. 
At high photon energies, the contributions of the two energy-dependent terms would approach zero, which explains why the {\it hit} and {\it digi} data would 
effectively be the same at high enough photon energies and given by the constant term $a$.
In order to check for systematic variations due to the edge effects, we choose a well-defined cluster (index {\it cluster}) inside the FwEndCap and compare the response in energy resolution (FWHM) and 
full-energy efficiency ($\epsilon$) with the average over the FwEndCap. To this end we simulate an isotropic photon distribution (index {\it average}) and define the following differences:
\begin{eqnarray}
\Delta_{res} &=& \left(\frac{FWHM}{2.35 \mu}[\%]\right)_{cluster} - \left(\frac{FWHM}{2.35 \mu}[\%]\right)_{average} \nonumber \\
\Delta_{eff} &=& {\epsilon}_{cluster} - {\epsilon}_{average}
\end{eqnarray}
for the differences in energy resolution and full-energy efficiency, respectively.
For various photon energies, Fig.~\ref{resolution_efficiency} shows the correlation between $\Delta_{res}$ and $\Delta_{eff}$.
The fact that 2~GeV photons have the largest change in efficiency and the smallest change in resolution confirms that the energy leakage from the FwEndCap edges is larger at 2~GeV than at the 
lower photon energies. At the lower photon energy of 100~MeV, the energy leakage is small and causes little change in the full-energy efficiency but the impact of edge leaking on the 
energy resolution is large.

\subsubsection{Light-yield non-uniformity}
\label{Light-yield-non-uniformity}
In the simulations of Fig.~\ref{FwEndCapPhotonResolution_1} and ~\ref{resolution_efficiency}, a uniform light transport is assumed as a function of the distance of the interaction 
point to the photo-sensor. In reality, however, non-uniformities may occur due, for instance, to the crystal geometry and one needs to validate the simulations by extracting the non-uniform light 
attenuation and light collection from prototype experiments. It has been shown that a stable temperature is of utmost importance in keeping a uniform detector response for PWO crystals
\cite{Grape}. Contributions to the non-uniformity for both tapered and non-tapered crystals originate from light attenuation along  the crystal due to 
intrinsic absorption inside the material, reflective properties of the crystal surface, transmission through the surface, the wrapping material as well as from diffusion on impurities and 
bubbles. In general, light collection of the full-size PWO crystal is a result of the interplay of optical absorption and focusing effect. Non-tapered crystals have the same value of the light yield along the 
length, because the focusing effect is negligible. 
\begin{figure}
\begin{center}
\resizebox{0.48\textwidth}{!}{\includegraphics{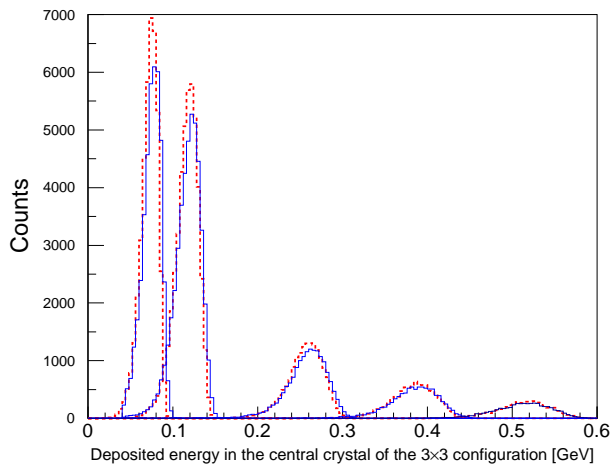}}
\caption{The superimposed spectra at five photon energies comparing simulations for a light yield of 500 photons/MeV, a zero noise level, and a 1.5\%/cm linear non-uniformity 
(red dashed-histogram) and experimental data (blue solid histogram). The spectra show the deposited energy in the {\it central} crystal of the $3\times3$ crystal configuration of PROTO60 
at photon energies of 93.97, 152.72, 339.67, 509.88, and 685.58~MeV. The absolute energy scale was normalized to the simulation at 685.58~MeV.}
\label{energy_centralCrystal}
\end{center}
\end{figure}
\begin{figure}
\begin{center}
\resizebox{0.48\textwidth}{!}{\includegraphics{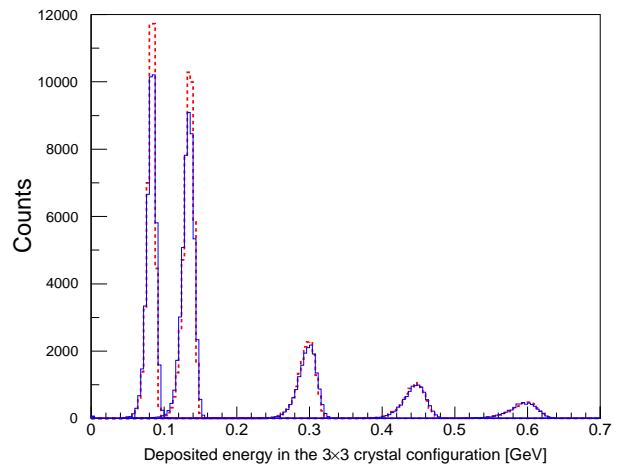}}
\caption{Same as Fig.~\ref{energy_centralCrystal}, but showing spectra of the deposited energy in the $3\times3$ crystal configuration of PROTO60.}
\label{sumEnergy9crystals}
\end{center}
\end{figure}
\par
\begin{figure*}
\begin{center}
\resizebox{0.95\textwidth}{!}{\includegraphics{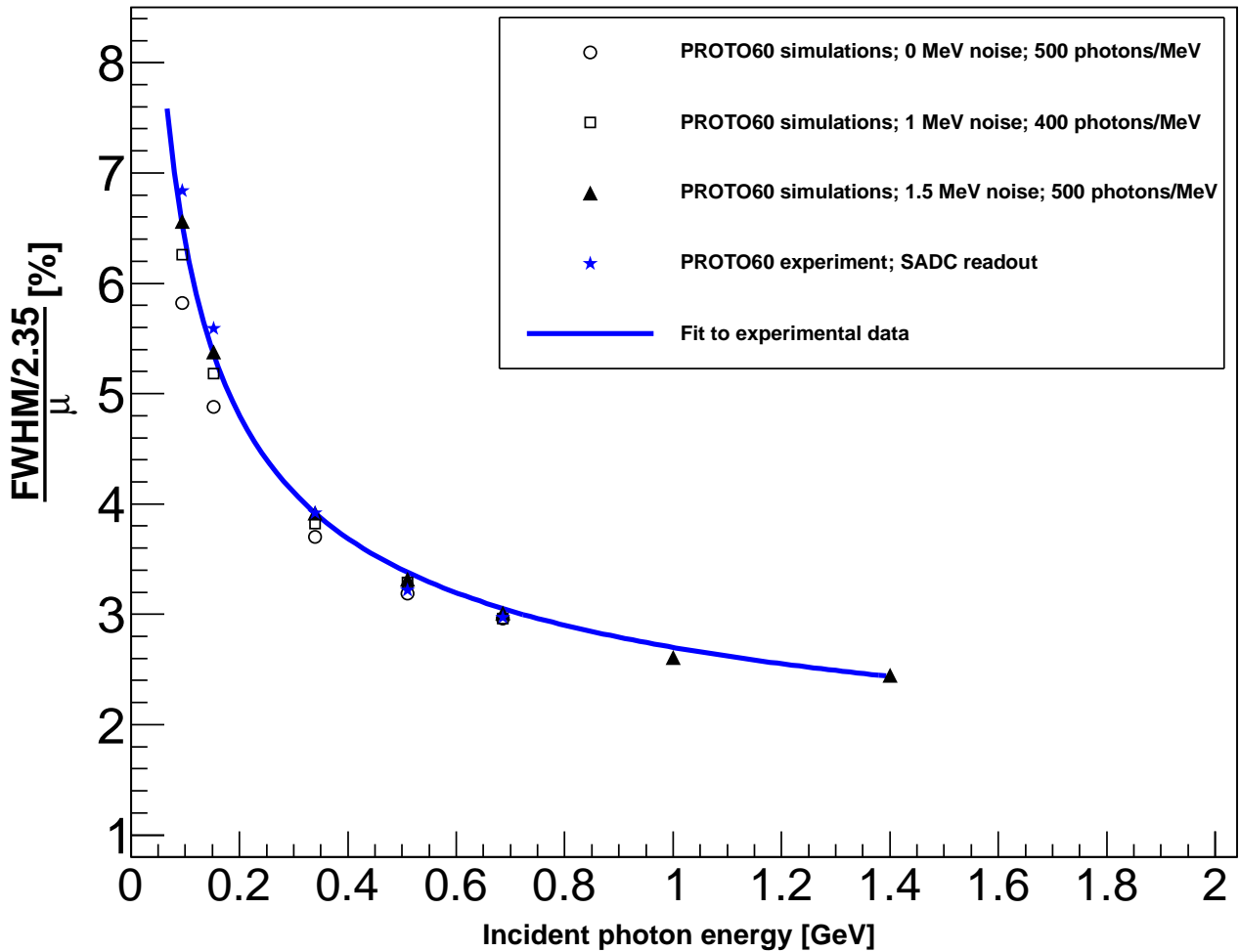}}
\caption{Simulation results (triangles, squares, and circles) for the energy resolution of the PROTO60 experiment (blue stars). The data and simulations are shown for the $3\times3$ crystal 
configuration with a 2~MeV detection threshold. The solid curve represents a fit to the data with the functional form of $1+\frac{1.7}{\sqrt{E}}$ \cite{Kavatsyuk}, where the constant 1 
arises from uncertainties in the cross calibration of the nine crystals. This function is in agreement with the fit according to $1.95\oplus \frac{1.9}{\sqrt{E}}$ within the error bars. Generally, a linear 
non-uniformity of 1.5\%/cm is introduced.
}
\label{proto60resolution}
\end{center}
\end{figure*}

 Based on the analysis of an array of $10\times6$ Barrel-type crystals in the PROTO60 experiment \cite{Kavatsyuk} and the comparisons with simulations, we could derive the 
non-uniformity from the data. In the simulations, the non-uniformity is implemented and calculated at the level of the interaction point where some energy is deposited. The 
procedure makes sure that only a specific percentage of this deposited energy will be taken into account depending on the distance of the interaction point to the face of the crystal. For 
instance, introducing an overall 30\% linear non-uniformity means that 100\%, 85\%, and 70\% of the deposited energy is collected, respectively, for an interaction point located at 
the front face, middle, and the back face of the crystal. The analyses were performed for a $3\times3$ crystal configuration in which every crystal was equipped with one LAAPD 
photo-sensor of $10\times10$~mm$^{2}$ area yielding a sensitivity of 52 photo-electrons/MeV. In the experiment, bremsstrahlung photons produced by an electron beam were collimated and focused 
at the center and perpendicular to the front face of the central crystal. The digitized data were obtained having a 2~MeV threshold and using the readout electronics based on the SADC. 
The values of the noise and non-uniformity parameters have been extracted by comparing simulations and data for a grid of values of non-uniformity and noise.
A non-uniformity in the light yield, decreasing linearly by 1.5\%/cm from the entrance to the photo-sensor face of the crystal, 
could explain the peak positions simultaneously in the energy-deposition spectra of the central crystal in a cluster as well as the energy sum of the nine crystals over a certain energy range. 
Figure~\ref{energy_centralCrystal} compares the superimposed spectra of the resulting simulations and the experimental data for five photon energies. The spectra show the deposited energy 
in the central crystal of the $3\times3$ crystal configuration of PROTO60. Figure~\ref{sumEnergy9crystals} compares, for the same photon energies, the simulations and experimental data for 
the deposited energy in the $3\times3$ crystal configuration of PROTO60.
The simulated spectra of Figs.~\ref{energy_centralCrystal} and~\ref{sumEnergy9crystals} contain the same number of events as the experimental spectra at the respective photon energy.
Figures~\ref{energy_centralCrystal} and \ref{sumEnergy9crystals} are obtained, assuming a light 
yield of 500 photons/MeV and a zero noise level in the simulations of the detector response. This low noise level is in 
accordance with the digital analysis of the PROTO60 data \cite{Kavatsyuk} and the parameterization of the energy resolution where the noise term is found to be absent \cite{Loehner}. 
Adopting these parameters, we compared the energy resolutions obtained in the simulations with the PROTO60 experiment. Figure~\ref{proto60resolution} shows the resolution as a function of 
energy for the $3\times3$ crystal configuration with a 2~MeV detection threshold for each crystal. 
Different simulations were performed for different values of the noise level and light yield, in order to check the 
sensitivity of the resolutions to these parameters. The optimized values for the noise level and light yield were obtained to be 1.5~MeV and 500 photons/MeV, respectively. These 
parameters together with the 1.5\%/cm linear non-uniformity will be used in Section~\ref{Charmonium_label} for the Barrel crystals to study the detector performance for a benchmark physics channel.
Experimental studies of the End-cap crystals have shown no indications of non-uniformity in the light yield of these crystals \cite{DBremer}.
\subsection{Cluster energy correction}
\label{FwEndCap_correction_label}
In the process of reconstructing and retrieving the correct energy and direction of the photons in the simulations, we used an optimized lookup table as a function of the cluster energy and polar angle. 
Cluster energies need to be corrected, in order to take into account the energy losses due to leakage and energy thresholds for single crystals. In order to construct the lookup table for the 
energy-angle correction of the clusters, simulations were performed with photons emitted isotropically from the target point at energies distributed uniformly between 2~MeV and 10~GeV. 
This energy range is large enough to include even the highest energy gammas expected in the study of the physics channel of interest that will be discussed in Section~\ref{Charmonium_label}.
A two-dimensional histogram was filled in $(E_{cl},\theta_{cl})$ bins for each cluster at a certain polar angle $\theta_{cl}$ and energy $E_{cl}$. The assigned 
energy $E_{cl}$ was taken from the predefined PandaROOT function for retrieving the deposited energy of the cluster without any correction. This function returns the total energy of 
{\it digi}s in the cluster. The weight for filling a certain 2D bin was defined by the ratio of the Monte Carlo energy of 
the generated photon and the retrieved energy of the cluster with highest deposited energy.
This was invoked to minimize the contribution of split-off clusters such that the direction of the reconstructed cluster is close to the direction of the photons generated in the Monte Carlo 
simulations. Finally, the energy correction value corresponding to each bin is taken to be the mean value of the bin content. This is also the case for the direction ($\theta$) correction 
which is taken as the difference between the Monte Carlo and reconstructed angle of the cluster with highest deposited energy.
In practice, the lookup table for correcting the cluster energies needs to incorporate the found non-uniformity of 1.5\%/cm for the Barrel crystals.
In the following we will make use of the obtained lookup table whenever we refer to cluster energies.
\begin{figure}
\begin{center}
\resizebox{0.48\textwidth}{!}{\includegraphics{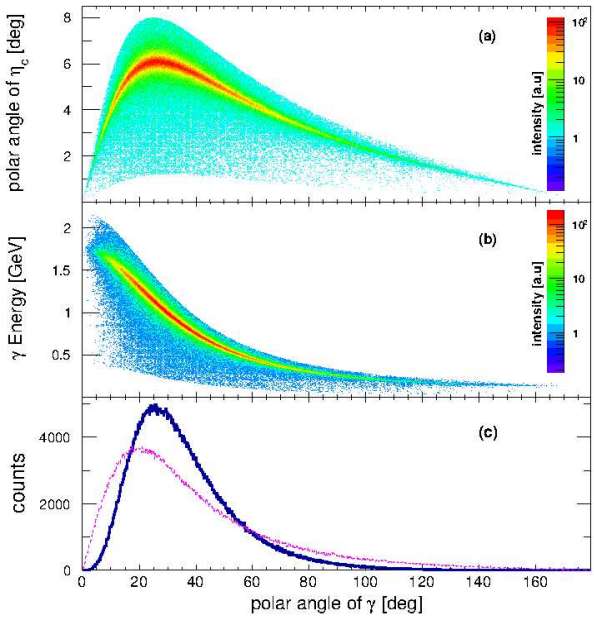}}
\caption{Monte Carlo simulations of kinematical variables in the laboratory system for $h_{c}$ production in $p\overline{p}$ annihilation and the decay channel 
$h_{c}\rightarrow \eta_{c}+\gamma$ assuming the $\sin^{2}(\theta_{CM})$ decay angular distribution from \cite{RadiativeAngularDistribution}. 
The beam energy was tuned to the production threshold energy of 6.6231~GeV.
a) polar scattering angle of $\eta_{c}$ as a function of the angle of the transition $\gamma$; b) energy of the transition $\gamma$ as a function of its scattering angle; c) solid blue histogram: 
distribution of the scattering angle of the transition $\gamma$; dashed magenta histogram: same for the 6 $\gamma$ generated in the $\pi^{0}$ and $\eta$ decays as well as the 
transition $\gamma$ when generated in the phase space decay of the $h_{c}$.}
\label{kinematics}
\end{center}
\end{figure}
\begin{figure}
\begin{center}
\resizebox{0.48\textwidth}{!}{\includegraphics{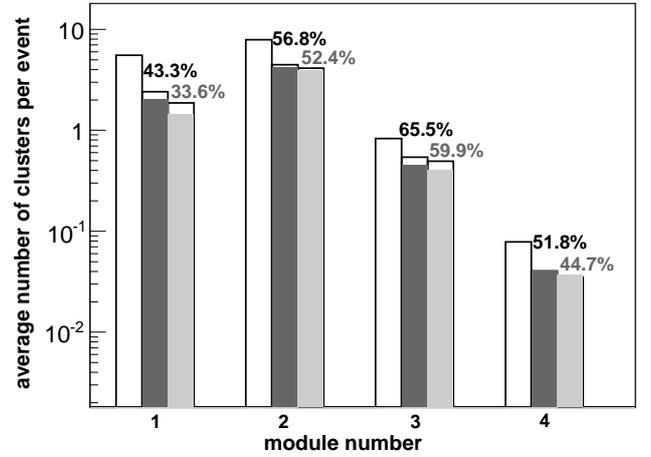}}
\caption{Simulation results for the average number of clusters per $h_{c}$ decay in the four EMC modules: 
FwEndCap (1), Forward Barrel (2), Backward Barrel (3), and BwEndCap (4) in the decay 
channel $h_{c}\rightarrow \eta_{c}\gamma\rightarrow \pi^{0}\pi^{0}\eta\gamma\rightarrow7\gamma$. 
The results are presented for cluster threshold energies of 3 (white), 10 (dark gray), and 30~MeV (light gray). 
The percentages of the registered clusters after applying a cluster threshold of 10~MeV (upper numbers) and 30~MeV (lower 
numbers), as compared to the 3~MeV threshold, are also shown above the corresponding bars in the graph. The shaded regions in each bar correspond to the respective percentages 
for 1~GeV photons emitted isotropically from the target point. In this case, the presented results are determined for the equivalent number of decayed $h_{c}$ resonances.}
\label{moduleDistribution}
\end{center}
\end{figure}
\section{Charmonium $h_{c}$ analyses}
\label{Charmonium_label}
\subsection{Cluster energy and multiplicity}
\label{Charmonium-cluster_label}
One of the goals of the $\overline{\rm{P}}$ANDA physics program is to measure the mass and the width of charmonium states with a resolution of $50-100$~keV. Charged reaction products 
together with photons are to be measured and analyzed. For the charmonium $h_{c}(1P)$ state, with the mass of $3525.31\pm 0.11{\rm (stat.)}\pm 0.14{\rm (syst.)}$~$\rm{MeV}/c^2$ and a width 
of $\Gamma=0.7\pm 0.28\pm 0.22$~$\rm{MeV}$~\cite{PRD86}, 
various decay modes can be investigated such as $J/\psi\pi\pi$, $\pi^{+}\pi^{-}\pi^{0}$, $2\pi^{+}2\pi^{-}\pi^{0}$, $3\pi^{+}3\pi^{-}\pi^{0}$, and $\eta_{c}\gamma$. Among these, the 
latter two have been observed~\cite{PDG}. In order to test the performance of the $\overline{\rm{P}}$ANDA calorimeter, 
a simulation study was performed for the representative decay channel $\overline{p}+{p} \rightarrow h_{c} \rightarrow \eta_{c}+\gamma \rightarrow (\pi^{0}+\pi^{0}+\eta)+\gamma \rightarrow 7\gamma$. 
Thus, in the final state, we would have 7 photons to be detected by the $\overline{\rm{P}}$ANDA detector and to be used to reconstruct the invariant mass of final-state mesons and the $h_{c}$ resonance. 
\par Figure~\ref{kinematics} shows Monte Carlo results for the relation between various kinematical variables of the two decay products of the $h_{c}$ in $p\overline{p}$ annihilation, 
namely the transition $\gamma$ and the $\eta_{c}$ assuming the $\sin^{2}(\theta_{CM})$ decay angular distribution from \cite{RadiativeAngularDistribution}. 
The mass of the $h_{c}$ was taken to be 3525.3~MeV/c$^2$, the width 0.7~MeV. The sensitivity of the kinematical variables to the width of the $h_{c}$, about 1~MeV, was found to be negligible.
The beam energy was tuned to the production-threshold energy of 6623.1~MeV. This situation simulates the beam-scanning technique \cite{PhysicsBook,PandaTechDesRep} of resonances and a beam-energy width 
of 0.55~MeV was assumed corresponding to the high-luminosity case of the HESR ($\frac{\sigma_{p}}{p}\approx10^{-4}$). For the decay width of the $\eta_{c}$, a value of 29.7~MeV was 
used taken from the PDG \cite{PDG}.
The angular distributions of the six decay gammas produced from the decay of $\eta$ and the two $\pi^{0}$ resemble the one for the transition $\gamma$ (dashed magenta line) in Fig.~\ref{kinematics}(c). 
The decay probability is peaked around $25^{\circ}$ and $6^{\circ}$ for the decay angles of the transition $\gamma$ and the $\eta_{c}$, respectively. 
Comparing Fig.~\ref{kinematics}(a) and (b), it is also interesting to note that the most forward-scattered 
transition $\gamma$ ($\eta_{c}$ particles) are the most (least) energetic, as the total energy of the transition $\gamma$ and $\eta_{c}$ should remain conserved.

\begin{figure*}
\begin{center}
\resizebox{0.98\textwidth}{!}{\includegraphics{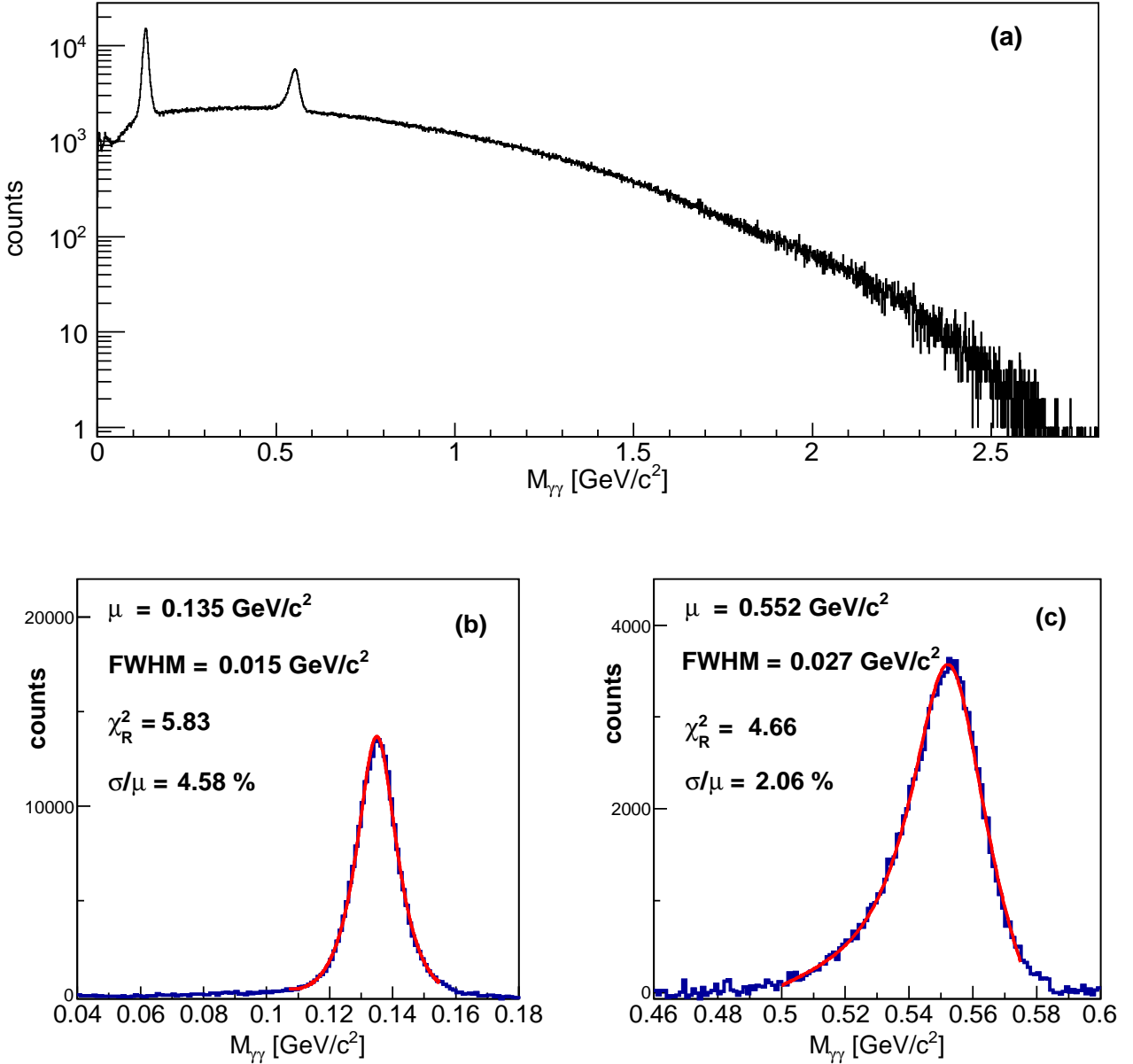}}
\caption{The invariant mass spectrum, reconstructed in the simulations of the full EMC, taking into account the contribution of all combinations of clusters that would possibly 
reproduce $\pi^{0}$ (b) and $\eta$ (c) particles before (a) and after (b and c) background subtraction. A threshold energy of 30~MeV is considered for all the clusters reconstructed by various EMC 
modules and a 1.5\%/cm non-uniformity is introduced only for the Barrel crystals. The 
fits are Lorentzian functions whose peak positions $\mu$ [GeV] and FWHM [GeV] are indicated in the lower panels.}
\label{invMass_pi0_eta}
\end{center}
\end{figure*}
\begin{figure*}
\begin{center}
\resizebox{0.98\textwidth}{!}{\includegraphics{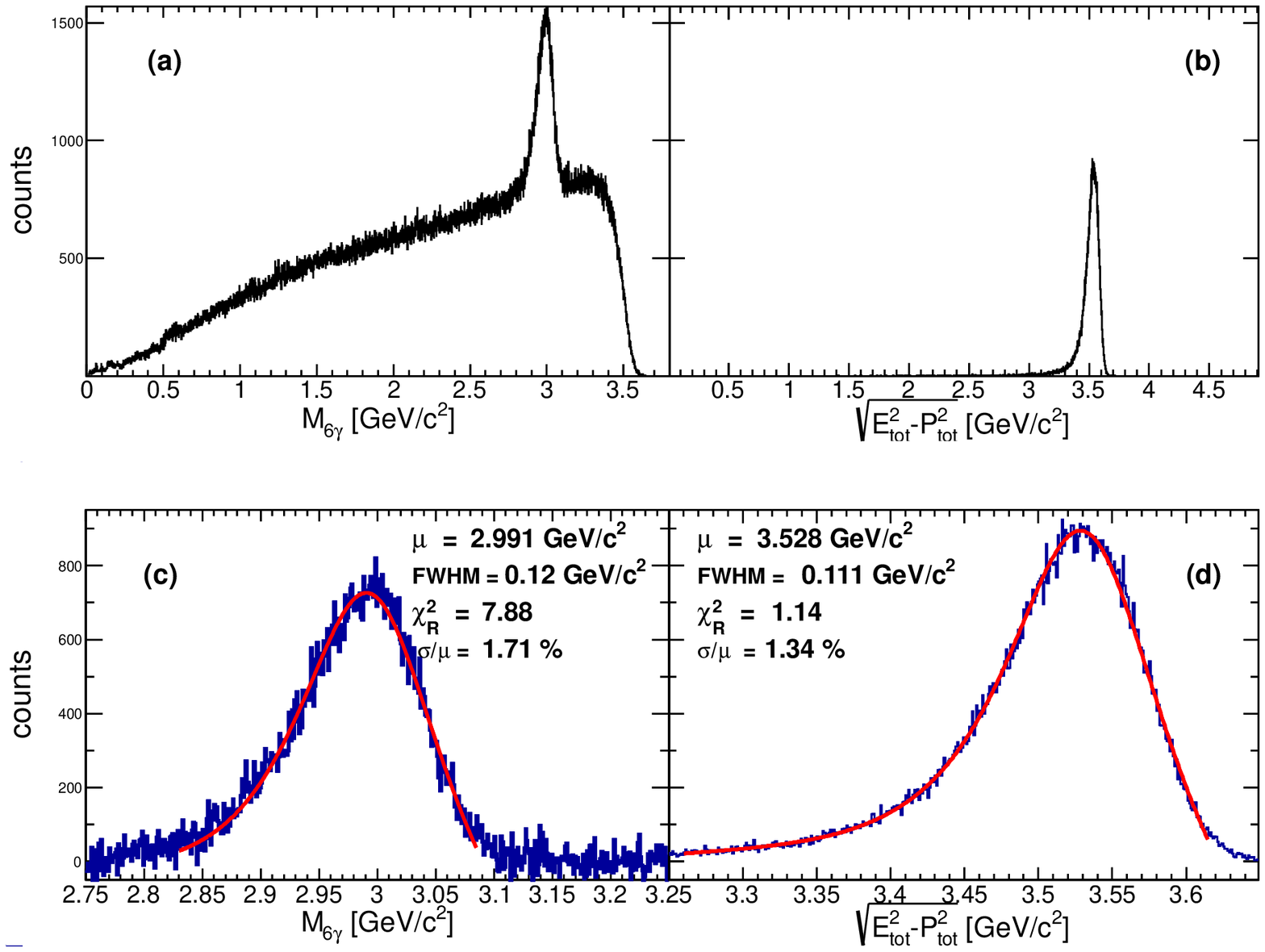}}
\caption{Left panels: the invariant mass spectra of $\eta_{c}$, reconstructed in the simulations of the full EMC, taking into account the contribution of all combinations of clusters 
that would possibly reproduce $\eta_{c}$ particles before (a) and after (c) background subtraction. The number of combinations per event for $\eta_{c}$ is $\binom{N_{cl}}{6}$, 
with $N_{cl}~(\geq7)$ being the number of clusters satisfying the cluster threshold energy of 30~MeV. In the case of $h_{c}$ (right panels), no background (b, expanded view in d) is assumed 
to be subtracted due to the way the $h_{c}$ spectrum is reconstructed: the invariant mass is calculated as the magnitude of the total four-momentum of all registered clusters for 
events with $N_{cl}\geq7$. The fits and parameters are defined as in Fig.~\ref{invMass_pi0_eta}.}
\label{invMass_eta_c_h_c}
\end{center}
\end{figure*}
\par The lookup table defined in Section~\ref{FwEndCap_correction_label} was applied for the analysis of 
the $h_{c}\rightarrow \eta_{c}+\gamma$ channel to correct the cluster energies retrieved with PandaROOT. 
Figure~\ref{moduleDistribution} shows the simulation results for the distribution of the four EMC modules registering clusters for the decay channel 
$h_{c}\rightarrow \eta_{c}\gamma\rightarrow \pi^{0}\pi^{0}\eta\gamma\rightarrow7\gamma$. The results are presented for cluster threshold energies of 3, 10, and 30~MeV. The 
percentages of the registered clusters after introducing cluster threshold energies of 10 and 30~MeV, as compared to the 3~MeV threshold, are also presented in this figure for 
different modules. The additional bias on the average number of clusters when we introduce a 30~MeV rather than a 10~MeV threshold is about 5\% for each EMC module except for the FwEndCap 
for which the average number of clusters shrinks by about 10\%.
\par If we perform the same simulations with seven uncorrelated photons of 1~GeV emitted isotropically from the target point, we get about the same results (the corresponding percentages are indicated by the shaded bars 
in Fig.~\ref{moduleDistribution}). It should be noted that the FwEndCap exhibits a larger reduction in the detected number of clusters than the other modules 
for a 10~MeV cluster threshold relative to the 3~MeV threshold. This effect is related to the previously mentioned overlap of 
the forward edge of the Barrel EMC with the FwEndCap circumference. 

\subsection{Signal invariant-mass reconstruction}
\label{Signal_label}
In order to reconstruct the invariant mass for pairs of registered clusters, one can make use of the following equation for the 2$\gamma$-decay of $\pi^{0}$ and $\eta$:
\begin{equation}
M_{\pi^{0},\eta}=\sqrt{2E_{1}E_{2}(1-\cos{\theta})},
\label{invMass_equation}
\end{equation}
where $M_{\pi^{0},\eta}$ is the invariant mass of the supposed $\pi^{0}$ or $\eta$ particles, while $E_{1}$, $E_{2}$, and $\theta$ represent the reconstructed energies and 
opening angle between the two clusters under analysis, respectively. In the simulations of the $h_{c}$ decay to $\eta_{c}\gamma\rightarrow \pi^{0}\pi^{0}\eta\gamma\rightarrow7\gamma$, we 
took all the possible combinations of choosing 2 photons out of the number of registered clusters satisfying the cluster threshold energy. The total number of cluster 
combinations is the binomial coefficient $\binom{N_{cl}}{2}$, in which $N_{cl}$ is the number of registered clusters per event. Thus, the invariant mass histogram is filled
$\binom{N_{cl}}{2}$ times per event for those events with $N_{cl}\geq7$. In this analysis we observed a maximum of 17, 33, and 67 registered clusters in the EMC, when we introduced 
cluster threshold energies of 30, 10, and 3~MeV, respectively. Figure~\ref{invMass_pi0_eta} shows the reconstructed invariant-mass spectra of $\pi^{0}$ and $\eta$ particles before 
and after subtracting the combinatorial background extracted from polynomial fitting, requiring at least 7 clusters to satisfy a cluster threshold energy of 30~MeV.
\par The invariant mass spectrum of $\pi^{0}/\eta$ can be employed to reconstruct the invariant masses of $\eta_{c}$ and $h_{c}$. In the 
case of $\eta_{c}$, the total number of cluster combinations per event is $\binom{N_{cl}}{6}$ for those events with $N_{cl}\geq7$. For each cluster 
combination, the invariant mass is calculated as the magnitude of the total four-momentum of the 6 clusters satisfying the cluster threshold. One can, in principle, follow 
the same procedure and add up the contribution of $\binom{N_{cl}}{7}$ combinations of clusters to get the invariant mass spectrum of $h_{c}$. However, we did obtain this spectrum 
by taking the magnitude of the total four-momentum of all registered clusters with total number of $N_{cl}\geq7$ per event. This way, we remove the huge amount of the combinatorial 
background that unavoidably enters in the cases of $\pi^{0}$, $\eta$, and $\eta_{c}$ due to the wrong combinations of clusters. Figure~\ref{invMass_eta_c_h_c} shows the 
reconstructed spectra of the invariant masses of the $h_{c}$ and $\eta_{c}$ (before and after background subtraction), assuming a 30~MeV cluster threshold.
\begin{table*}[ht!]
\caption{Fitting parameters for the spectra of the invariant masses and total cluster energies as well as derived resolutions (FWHM/2.35) in detecting $\pi^{0}$, $\eta$, 
$\eta_{c}$, and $h_{c}$ particles in the $h_{c}$ decay channel, assuming a cluster threshold energy of 30 (10)~MeV. All spectra were obtained by taking into account the 
contribution of only those events with at least 7 clusters satisfying the threshold. The results are shown for the three cases of having the full EMC, or only the FwEndCap, or the 
full $\overline{\rm{P}}$ANDA detector present in the analysis. Clearly, the geometrical acceptance of the FwEndCap does not allow detecting simultaneously the 6 or 7 gammas emitted 
from the decay of $\eta_{c}$ or $h_{c}$. For the error calculation, the quality of fitting has been taken into account through multiplying the error boundaries by $\sqrt{\chi^{2}_{R}}$ to 
obtain the indicated error boundaries.}
\begin{center}
\begin{tabular}{ | c | c | c | c | c |} \hline
\multicolumn{1}{|c}{decayed particle} & 
\multicolumn{1}{|c}{$\mu$ [MeV]}            & 
\multicolumn{1}{|c}{FWHM/2.35 [MeV]}       &
\multicolumn{1}{|c}{$\chi^{2}_{R}$}       &
\multicolumn{1}{|c|}{resolution [\%]}      \\ \hline
\multicolumn{5}{|c|}{Full EMC}         \\ \hline

            & $134.98\pm0.02$    & $6.18\pm0.05$     & 5.83     & $4.58\pm0.1$                  \\
 $\pi^{0}$  &                        &                         &          &                                \\
            & ($134.74\pm0.02$)  & ($6.11\pm0.04$)   & (6.18)   & ($4.53\pm0.07$)                  \\ \hline
            & $552.12\pm0.07$    & $11.38\pm0.18$     & 4.66     & $2.06\pm0.06$                    \\
 $\eta$     &                        &                         &          &                                \\
            & ($552.3\pm0.09$)  & ($11.35\pm0.25$)   & (6.92)    & ($2.05\pm0.13$)       \\ \hline
            & $2990.6\pm0.4$     &  $51.01\pm0.72$    &  7.88    & $1.71^{+0.06}_{-0.08}$         \\
 $\eta_{c}$ &                        &                         &          &                       \\
            & ($2993.1\pm0.38$)   & ($50.12\pm0.83$)   & (7.64)   & ($1.67^{+0.08}_{-0.05}$)              \\ \hline
            & $3528.44\pm0.36$    & $47.42\pm0.72$     & 1.14     & $1.34\pm0.02$        \\
 $h_{c}$    &                        &                         &          &                                \\
            & ($3530.5\pm0.38$)  & ($46.54\pm0.63$)   & (1.23)   & ($1.32\pm0.02$)      \\ \hline
            & $6650.78\pm0.53$    & $73.7\pm1.05$     & 1.04     & $1.11^{+0.01}_{-0.02}$                    \\
 $h_{c}$    &                        &                         &          &                                \\
total energy& ($6661.08\pm0.43$)  & ($72.71\pm0.89$)   & (1.35)   & ($1.09\pm0.01$)      \\  \hline
\multicolumn{5}{|c|}{FwEndCap}         \\ \hline

            & $137.68\pm0.05$    & $5.81\pm0.12$     & 3.76     & $4.22\pm0.17$        \\
 $\pi^{0}$  &                        &                         &          &                                \\
            & ($137.45\pm0.06$)  & ($5.59\pm0.11$)   & (4.86)   & ($4.07^{+0.18}_{-0.2}$)      \\ \hline
            & $554.25\pm0.36$    & $10.12\pm0.82$     & 3.71     & $1.82^{+0.29}_{-0.27}$        \\
 $\eta$     &                        &                         &          &                                \\
            & ($554.71\pm0.46$)  & ($10.88\pm0.75$)   & (4.23)   & ($1.96\pm0.29$)      \\ \hline
\multicolumn{5}{|c|}{}         \\

\multicolumn{5}{|c|}{Full $\overline{\rm{P}}$ANDA Detector}         \\ \hline

  $\pi^{0}$  & $134.76\pm0.03$    & $6.34\pm0.06$     & 5.6    & $4.70^{+0.12}_{-0.09}$                  \\ \hline
  $\eta$     & $551.97\pm0.05$    & $12.43\pm0.14$     & 4.2   & $2.25^{+0.06}_{-0.04}$                  \\ \hline
  $\eta_{c}$ & $2981.05\pm0.41$    & $53.06\pm0.19$     & 6.17   & $1.78\pm0.02$                   \\ \hline
  $h_{c}$    & $3520.32\pm0.56$    & $48.23\pm0.78$     & 2.14   & $1.37\pm0.03$         \\ \hline

\end{tabular}
\label{resolutions}
\end{center}
\end{table*}
\par Table~\ref{resolutions} summarizes the fitting parameters for the spectra of the invariant masses and total energy of all clusters, together with the derived resolutions in 
detecting $\pi^{0}$, $\eta$, $\eta_{c}$, and $h_{c}$ particles, assuming cluster threshold energies of 10 and 30~MeV. The resolutions are provided for the three cases of having the 
full EMC, or only the FwEndCap, or the full $\overline{\rm{P}}$ANDA detector (including magnets, pipe, STT, MVD, MTD, GEM, DIRC, DSK, FTS, FTOF, SciTil) present in the analysis. It is 
interesting to note that, for $\pi^{0}$ and arguably for $\eta$, the resolution results derived from the FwEndCap are slightly better than the ones extracted from 
the full EMC analysis. 
This is due to the boost in the forward angles, where the two high-energy photons detected by the FwEndCap would ensure a relatively better resolution than when they would be detected 
at larger opening angles by different EMC modules. The mean values of the $\pi^{0}$ and $\eta$ invariant masses reconstructed by the FwEndCap, as the only EMC module present in the geometry, are 
overestimated as compared with the ones obtained by the full EMC. This is due to the shower leakage from the corner edges of the FwEndCap and the performance of the lookup table for these regions. 
Clearly, this would not be the case for the full EMC geometry as, seen from the target point, the corner edges of the FwEndCap are covered by the Barrel forward edges.
\begin{table*}
\caption{Parameters and design performance of the EMC in various experiments compared with the ones of the $\overline{\rm{P}}$ANDA EMC. The results for the $\overline{\rm{P}}$ANDA 
and BESIII \cite{BESIII_NIM} calorimeters are from the Monte Carlo simulations.}
\begin{center}
\begin{tabular}{ | c | c | c | c | c | c | } \hline
  Parameter                           &  $\overline{\rm{P}}$ANDA  & BESIII      &  CLEO-c~\cite{CLEOEXP}     &  BaBar~\cite{BaBarExp}      &  Belle~\cite{BelleExp} \\ \hline
  Radiation length ($X_{0}$)          &  22                       & 15          &  16         &  $16-17.5$  &  16.2  \\ \hline
  $\sigma_{E}$ [MeV]                  &  25.4                     & $\approx25$ & $\approx20$ & $\approx28$ &  $\approx17$  \\
  at 1~GeV                            &                           &             &             &             &    \\  \hline
  $\sigma_{E}$ [MeV]                  &  6.6                      & 3.3         &  4          &  4.5        &  4  \\
  at 100~MeV                          &                           &             &             &             &    \\  \hline
  Position resolution ($\sigma$) [mm] &  3.3                      &  6          &  4          &  4          &  6  \\ 
  at 1~GeV                            &                           &             &             &             &     \\ \hline
\end{tabular}
\label{comparison_otherExperiments}
\end{center}
\end{table*}
\section{Discussion of simulation results}
\label{Summary_label}
Simulations were performed with photons emitted isotropically from the target as well as hitting specific crystals inside the FwEndCap.
Subsequently, various analyses were performed including the full-energy efficiency and the energy and position resolutions.
For 1~GeV isotropically-emitted photons, the position resolution of $\sigma_{x,y}\approx3.3$~mm for cluster reconstruction was obtained, which can be compared with the 
front-face extension of the crystal (24.37~mm). An investigation of the energy-weighting parameters resulted in best position resolutions for the FwEndCap, once we 
introduced:\\ $W_0=4.071-0.678\times \left(\frac{E_{cl}}{{\rm GeV}}\right)^{-0.534}\cdot \exp\left(-\left(\frac{E_{cl}}{{\rm GeV}}\right)^{1.171}\right)$.
\par The multiplicity distribution after digitization showed a significant sensitivity to the detection threshold of individual crystals.
As a result of the partial overlap of the FwEndCap circumference with the forward edge of the Barrel EMC, a shift to a lower mean {\it digi} multiplicity is observed. 
This effect can be considerable for high photon energies at high detection thresholds. 
The cluster-multiplicity distribution reveals as well a drastic sensitivity to the cluster threshold as expected; for 1~GeV photons and assuming a 30~MeV (10~MeV) cluster threshold, 
one would expect 99.6\% (97\%) of events to have a maximum of two registered clusters.
In principle, however, one has to be careful when implementing cluster thresholds in the analysis, e.g. when one needs to achieve $\pi^{0}$ background suppression based on the analysis of low energy photons.
Although we chose, in this paper, to suppress the split-off clusters by optimizing the cluster threshold, there is a need for more refined algorithms exploiting the threshold dependence.
The investigation of the full-energy efficiency of the FwEndCap as the only module in the geometry showed that, for a 3~MeV {\it digi} threshold, one can expect a collection efficiency of about 95\% for the 
photon energies ranging from 1~GeV to 10~GeV. The efficiency would decrease to 93\% as the photon energy decreases from 1~GeV to 100~MeV. Although the efficiency shows considerable sensitivity 
to the detection threshold at low photon energies, it would only change slightly (0.5\%) for high energy photons.
\par A comparison of the energy resolutions from simulations and the PROTO60 experiment with a sensitivity of 52 photo-electrons/MeV and 
2~MeV {\it digi} threshold shows very good agreement with less than 0.5\% worse resolution ($\sigma/\mu$) of PROTO60 at low energies of the order of 100~MeV.
We verified that the energy resolution for the Barrel-type crystals, in the energy range from about 100~MeV to 1.4~GeV, satisfies the functional form $\sigma/\mu[\%] = 1.95\oplus \frac{1.9}{\sqrt{E/\rm{GeV}}}$ which is close to 
the requirement stated in \cite{PandaTechDesRep} i.e. $\sigma/E[\%] \leq 1 \oplus \frac{\leq 2}{\sqrt{E/\rm{GeV}}}$.
\par For isotropic photon emission from the target and by increasing the cluster threshold from 3~MeV to 10~MeV, the number of clusters in each of the four modules of the EMC would 
diminish by about 50\%. The reason for a slightly stronger reduction of clusters in the FwEndCap lies in the partial overlap with and screening by the Barrel EMC. 
\par The simulation studies for the FwEndCap EMC were completed by analysing a representative physics channel. The charmonium $h_{c}$ decay into $\rm{\eta_{c}}\gamma$, assuming a decay 
probability proportional to $\sin^{2}(\theta_{CM})$, and the subsequent decay into $\pi^{0}\pi^{0}\eta\gamma$ were implemented in the event generator with $\pi^{0}$ and $\eta$ decaying into 2 photons. 
The analysis of this channel was conducted for the two cases of having the FwEndCap as the only EMC module in the analysis as well as for the full EMC. 
The constructed lookup table for energy-direction corrections of the clusters as a function of their energy and polar angle allowed to reconstruct invariant-mass peak positions and resolutions 
of the $h_{c}$ and its various decay products $\pi^{0}$, $\eta$, and $\eta_{c}$.
\begin{sloppypar}
The parameters and expected performance of the $\overline{\rm{P}}$ANDA EMC, based on prototype data and Monte Carlo simulations, can be compared with the corresponding ones of other 
(running) experiments, as presented in Table~\ref{comparison_otherExperiments}.
The EMC calorimeters in CLEO-c, BaBar, Belle, and BESIII detectors were all made of CsI(Tl) crystals. The design energy resolution of EM showers in BESIII is about 2.5\% and 3.3\% for 
photon energies of 1~GeV and 100~MeV, respectively. The corresponding resolutions for the Barrel EMC of $\overline{\rm{P}}$ANDA in presence (absence) of non-uniformity were obtained to be 
2.5\%~$\mathrel{\widehat{=}} 25.4$~MeV (1.3\%~$\mathrel{\widehat{=}} 12.7$~MeV) and 6.6\%~$\mathrel{\widehat{=}} 6.6$~MeV (5.5\%~$\mathrel{\widehat{=}} 5.6$~MeV), respectively.
The Monte Carlo result for the mass resolution of the $\pi^{0}$ reconstructed by the BESIII EMC in the decay channel of $J/\psi \rightarrow \rho\pi$ with $\pi^{0} \rightarrow \gamma\gamma$ has been 
determined to be $\sigma_{M}=7.3$~MeV \cite{BESIII_NIM}. We obtained a slightly better mass resolution of $\sigma_{M}=6.2$~MeV for the $\pi^{0}$ in the decay channel 
of $h_{c}$ to $\pi^{0}\pi^{0}\eta\gamma$ using the $\overline{\rm{P}}$ANDA EMC.
\end{sloppypar}

\section{Conclusions}
The mechanical design of the Forward End-cap EMC is ready for construction. 
Based on the defined geometry, the PbWO$_4$ crystals and the Carbon-fiber containers (alveoles) were implemented in the PandaROOT simulation package.
Simulations have been validated by prototype experiments.
The obtained energy resolutions of the Barrel-type crystals are satisfactory and close to the expectations raised by the technical design report. 
The overlap between the Barrel and the Forward End-cap EMC causes modifications of the observed cluster distributions. 
The optimized analysis of this overlap region requires developing a dedicated reconstruction algorithm. 
The analysis presented here shows that the overlap effects are of minor importance for the overall performance of the electromagnetic calorimeter.

\section{Acknowledgements}
\label{Acknowledgements_label}
\begin{sloppypar}
The authors thank Prof. N. Kalantar-Nayestanaki for useful comments. 
The authors also thank the EMC and computing groups of the $\overline{\rm{P}}$ANDA collaboration for their useful comments and suggestions.
This work was supported by the GSI Helmholtzzentrum f\"ur Schwerionenforschung GmbH, Germany, 
the University of Groningen, and the Dutch Organization for Scientific Research (NWO).
\end{sloppypar}
%

\end{document}